\documentclass{article}

\usepackage{graphicx} 
\usepackage[hidelinks]{hyperref} 
\usepackage{natbib} 
\usepackage{subcaption} 
\usepackage{enumitem} 
\usepackage{pifont} 
\usepackage{comment}
\usepackage[normalem]{ulem} 
 
\usepackage[skip=10pt plus1pt, indent=0pt]{parskip} 
\usepackage[margin=1in]{geometry} 

\usepackage{pdflscape}
\usepackage{booktabs} 

\usepackage{amssymb,amsmath,amsthm,amsfonts}
\usepackage{mathtools}
\usepackage{bbm}
\usepackage{color,xcolor}

\usepackage{enumitem}

\usepackage{multirow}

\usepackage{comment}

\usepackage{setspace}
\newcommand{\one}{\mathbbm{1}}
\newcommand{\E}{\operatorname{\mathbbm{E}}}
\newcommand{\Q}{\operatorname{\mathbbm{P}}}

\newcommand{\Xvec}{\boldsymbol{X}}

\theoremstyle{plain}

\theoremstyle{definition}

\theoremstyle{remark}

\newcommand{\specificthanks}[1]{\@fnsymbol{#1}}

\usepackage{authblk}

\title{\hrule Probabilistic AI models outperform numerical weather prediction models on record breaking heat extremes}
\author[1,\thanks{Correspondence: jakob.wessel@uni-leipzig.de}]{Jakob Benjamin Wessel}
\author[1,2]{Fiona Spuler}
\author[3]{Christopher A. T. Ferro}
\author[1]{Marlene Kretschmer}
\author[1]{Sebastian Sippel}
\affil[1]{Institute for Meteorology, Leipzig University, Germany}
\affil[2]{Center for Scalable Data Analytics and Artificial Intelligence, Leipzig University, Germany}
\affil[3]{Department of Mathematics and Statistics, University of Exeter, UK}

\date{\today\vspace{10pt}\hrule}

\begin{document}

\maketitle

\begin{abstract}

AI weather prediction (AIWP) models have emerged as a key complement to physics-based numerical weather prediction (NWP) models, and now often outperform them on aggregate skill. Yet, several studies report that physics-based models still forecast extreme events better, in particular record-breaking extremes, based on the evaluation of deterministic forecasts. Here we evaluate heat extreme forecasts from probabilistic AIWP models over global land and show that they clearly outperform the ECMWF IFS ensemble, both raw and post-processed, including for record-breaking events, in all regions and at all lead times considered. AIWP models better forecast the occurrence and intensity of extremes, are better calibrated in the tail and miss fewer records. Station observations confirm these results. We argue that many reported AIWP limitations stem from unsuitable verification choices, specifically comparing deterministic forecasts that target different quantities. Instead, a fair evaluation of extremes necessitates probabilistic verification.

\end{abstract}

\section{Introduction}
Numerical weather prediction (NWP) has seen a revolution in recent years with the emergence of AI weather prediction (AIWP) models. These models show impressive performance across a range of variables and metrics and frequently surpass classical physics-based NWP models in terms of standard and average verification scores such as the Root mean square error (RMSE) or Continuous ranked probability skill score (CRPS) \citep[e.g.][]{keisler_forecasting_2022, lam_graphcast_2022, bi_pangu-weather_2022, lang_aifs_2024}. AIWP models are based on deep neural network architectures such as graph neural networks \citep{lam_graphcast_2022}, optimized against a score. They are extremely fast and can produce new forecasts in minutes on a single graphical processing unit (GPU), rather than requiring the high-performance computing resources of a classical NWP system. Although most AIWP models still rely on initial conditions from physical data assimilation, the first models that forecast directly from observations have been emerging \citep{mcnally_data_2024, alexe_graphdop_2024, allen_end--end_2025}.

Aggregate skill as reported in benchmarks, however, is only part of what is required of a forecasting system. Weather forecasts underpin early warning systems, and extreme events carry outsized societal impacts, so the ability of AIWP models to predict extreme events is crucial. Accordingly, a growing literature has started to evaluate the performance of AIWP models in forecasting extreme events, finding that they often perform worse than physics-based NWP models. A number of studies have considered case studies of individual events \citep[e.g.][]{charlton-perez_ai_2024, pasche_validating_2025, afargan-gerstman_ai_2025, lopez-marti_can_2025, hewitt_evaluating_2026, nelli_june_2026, dacre_northern_2026}, reporting limitations of AIWP models, particularly in predicting the intensity of extreme events such as maximum wind speed of mid-latitude storms \citep{charlton-perez_ai_2024}. While such case studies provide valuable process-level insight into model behaviour, their conclusions might not readily generalise beyond the individual events and event types considered. To address this, \citet{zhang_physics-based_2026} recently conducted a global systematic statistical evaluation of AIWP and NWP forecasts of record-breaking temperature and wind extremes, finding that AIWP models underestimate both occurrence and intensity of record-breaking extremes, particularly record-breaking heat extremes, relative to classical physics-based NWP models.

\begin{figure}[t]
    \centering
    \includegraphics[width=0.9\linewidth]{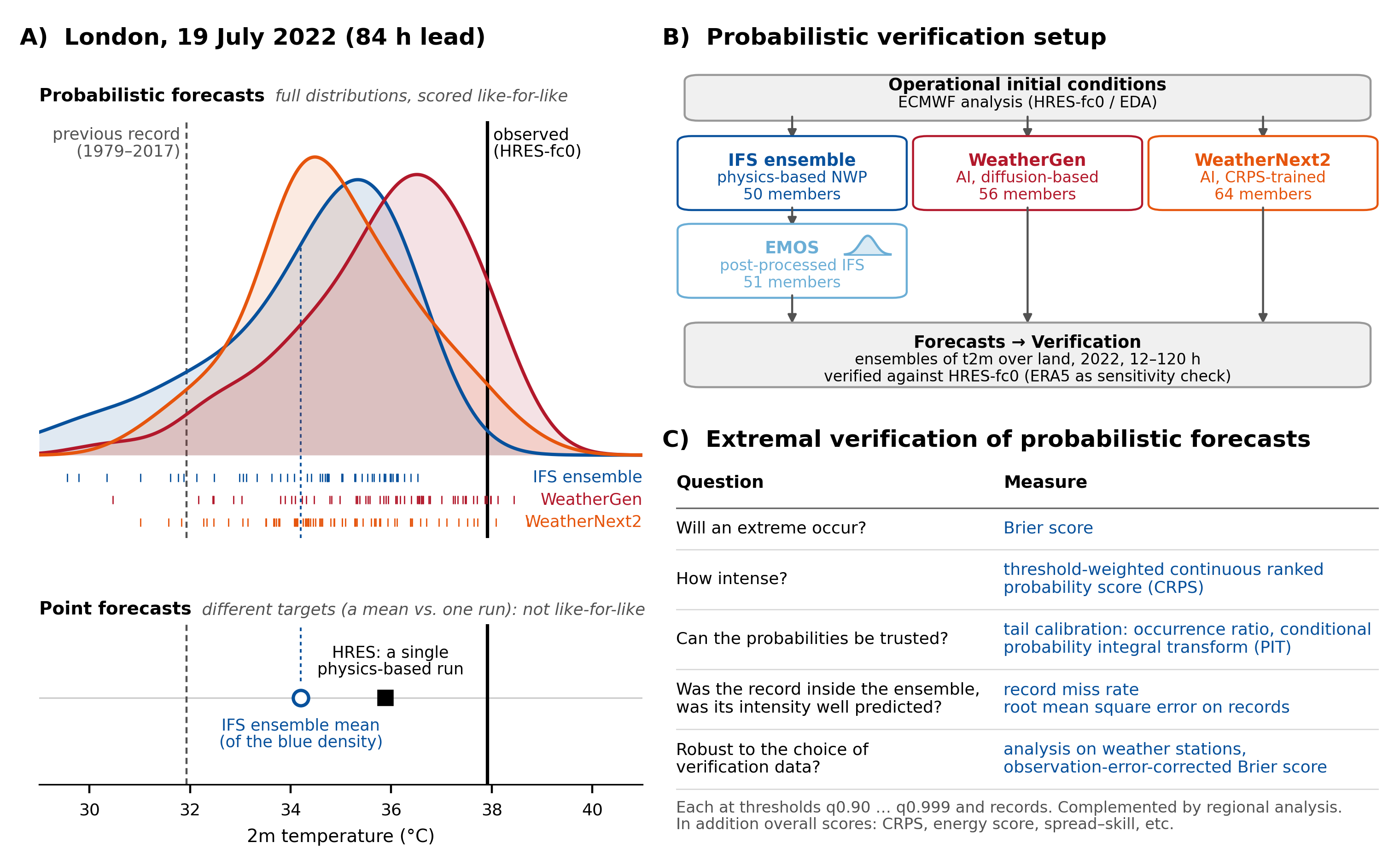}
    \caption{\textbf{Overview of the forecast evaluation protocol used in this study.} A) Probabilistic and deterministic forecasts of 2\,m temperature (t2m) in London, 19 July 2022 12\,UTC, 84\,h lead time. Top: predictive distributions (kernel density estimates; ticks mark members) of the physics-based IFS ensemble and the AI weather prediction (AIWP) models WeatherGen and WeatherNext2; all IFS members lie below the observation. Bottom: point forecasts from the IFS ensemble mean, an estimate of the conditional mean lying below the estimate from HRES, a single high-resolution physics-based run. Deterministic AIWP models trained on mean squared error also estimate the conditional mean, so comparing them to HRES is not like-for-like. B) Verification setup: models, IFS post-processing with ensemble model output statistics (EMOS), and verification data. C) Extremal scores and metrics considered and their intuitive interpretation.} 
    \label{fig:concept}
\label{fig:concept}
\end{figure}

All of these previous studies, including the systematic evaluation by \citet{zhang_physics-based_2026}, predominantly evaluate \textit{deterministic} AIWP models, and often benchmark them against HRES, the high-resolution run of the ECMWF-IFS. We argue that this comparison is not like-for-like, nor does it correspond to what operational forecasting centres are necessarily targeting. When issuing a deterministic (point) forecast of an uncertain quantity, it is critical to consider what the forecast represents. Deterministic AIWP models are trained by minimizing a (root) mean squared error: a loss that \emph{elicits} the mean of the predictive distribution \citep{gneiting_making_2011}: their output is thus a statistical estimate of the conditional mean. This is just as training with a quantile loss would yield an estimate of a conditional quantile. More generally, in the forecasting literature point forecasts are understood as forecasts of functionals of the underlying probability distribution of the outcome \citep{gneiting_making_2011, kolassa_why_2020, petropoulos_forecasting_2022}. Care must therefore be taken when evaluating deterministic (point) forecasts: evaluation metrics should be consistent with the functional that the forecast represents, and forecasts should only be compared against one another if they target the same functional. 

While the output of a deterministic AIWP model is a statistical estimate of the conditional mean,  HRES, in contrast corresponds to a single physical integration from the control member of the ensemble data assimilation. This means that AIWP models and HRES correspond to different forecast targets. HRES, as the "most likely physical integration", may realise, overshoot, or miss an extreme. For the AIWP models, however, smoothing towards the mean, and hence a muted representation of tail events, follows directly from the forecast target (the conditional mean) and training loss rather than indicating a deficiency of the model. As a consequence, their output cannot be interpreted as a physically consistent realisation of a possible atmospheric state in the way that HRES can. Comparing the two thus conflates the type of forecast with its quality, and reported deficits of AIWP models for extremes may therefore reflect the evaluation design, and the interpretation of what a deterministic forecast represents, as much as the models themselves. Indeed recent studies of optimal (in-sample) post-processing of deterministic AIWP models find that they carry more, not less, information about extreme events compared to their NWP counterpart \citep{gneiting_probabilistic_2026, biegert_towards_2026}. 

The ambiguities involved in comparing a single ensemble member of an NWP forecast with a deterministic AI forecast, estimating the mean of a predictive distribution, can be avoided altogether by evaluating probabilistic forecasts, the operational standard at most meteorological services (Fig.~\ref{fig:concept} provides an example of this for the July 2022 heatwave). Indeed, fully probabilistic AIWP models have recently become available, allowing for a direct comparison against NWP ensemble forecasts. A mature statistical verification methodology exists for probabilistic forecasts, from proper scoring rules \citep{gneiting_strictly_2007, waghmare_proper_2025} and their weighted versions \citep{gneiting_comparing_2011, lerch_forecasters_2017, allen_weighted_2023} to (tail) calibration diagnostics \citep{gneiting_probabilistic_2007, gneiting_regression_2023, allen_tail_2025}, allowing to evaluate both the information content present in the probabilistic forecast, and whether the derived probability statements are trustworthy. This set of verification metrics enables a comparison in which AIWP and NWP systems issue the same type of forecast and are held to the same standard.
 
In this work, we carry out a systematic evaluation of probabilistic AIWP and NWP forecasts for heat extremes and record-breaking heat extremes over global land areas (Fig.~\ref{fig:concept} provides an overview of our evaluation protocol). We evaluate two operational, state-of-the-art probabilistic AIWP models: WeatherGen / GenCast \citep[a diffusion-based model,][]{price_probabilistic_2025} and Weathernext2 / FGN \citep[a CRPS-trained model][]{alet_skillful_2025}, and benchmark them against the ECMWF-IFS ensemble. We focus on evaluating a single year, 2022, for which forecast data from all ensembles was available. For most meteorological centres, the raw model is not the final operational output, but instead it undergoes substantial post-processing \citep{roberts_improver_2023}, which includes statistical adjustments. Thus, in order to be closer to operational practice we also include the IFS ensemble adjusted with Ensemble Model Output Statistics (EMOS), a widely-used post-processing technique \citep{gneiting_weather_2005, vannitsem_2018_chapter_3}.

All forecasts are initialized from operational ECMWF analyses and verified against the HRES initial conditions (\texttt{HRES-fc0}), following standard practice in AIWP evaluation \citep[e.g.,][]{bi_pangu-weather_2022, rasp_weatherbench_2024}. We define heat extremes as exceedances of grid-cell-, month-, and forecast-hour-specific quantiles derived from the 1979--2017 ERA5 reanalysis climatology, and record-breaking heat extremes, or records, as exceedances of the historical maximum over this baseline period. Crucially, because these thresholds are fixed \emph{a priori} using historical climatology, our evaluation avoids the statistical artefacts associated with conditioning on observed outcomes \citep{lerch_forecasters_2017}, and matching thresholds by month means that both Summer and Winter heatwaves are included. We assess probabilistic forecasts (including for extreme events) adhering to the principle that forecasts should be maximally \emph{sharp} i.e. informative and \emph{calibrated}, so users can trust the probabilities derived from the forecast \citep{gneiting_probabilistic_2007}. We assess both jointly using different proper scoring rules, including for heat extremes the Brier score \citep{brier_verification_1950} to assess occurrence forecasts, and the threshold-weighted CRPS \citep{gneiting_comparing_2011} to assess occurrence and intensity forecasts jointly. We evaluate whether predicted probabilities for heat extremes can be trusted using (tail) calibration diagnostics \citep{vannitsem_2018_chapter_6, allen_tail_2025} and metrics on forecast miss rates. Finally analysis will be complemented by regional perspectives and analysis on the sensitivity towards verification data.

\section{Results}

\subsection{AIWP models outperform the IFS ensemble in all proper scores and at all lead times}

\begin{figure}[t]
    \centering
    \includegraphics[width=1.\linewidth]{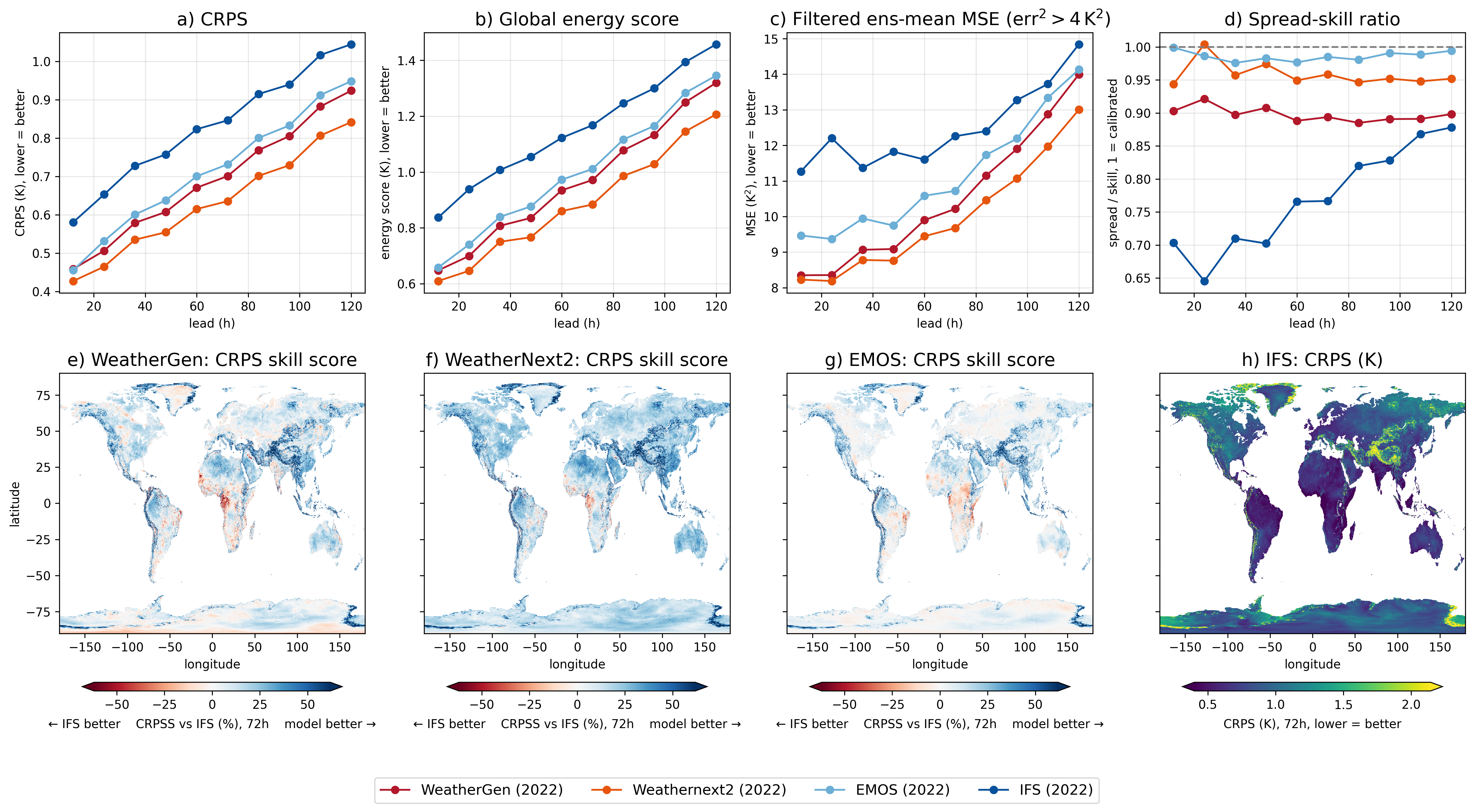}
    \caption{\textbf{Overall performance of the AIWP models (WeatherGen, WeatherNext2), the raw IFS ensemble and EMOS-post-processed IFS for 2m temperature over land in 2022, verified against HRES-fc0.} Top row: probabilistic scores as a function of lead time, a) CRPS, b) energy score, c) mean squared error of the ensemble mean restricted to large errors (squared error $>4\,\mathrm{K}^2$), d) spread-skill ratio (dashed line at 1 indicates a calibrated forecast). Lower is better for a)--c). Bottom row: maps at 72h lead time of e)--g) the CRPS skill score (\%) of WeatherGen, WeatherNext2 and EMOS relative to the IFS (positive values mean the model beats the IFS) and h) the CRPS (K) of the IFS. Figure \ref{fig:app_scores_overview} in the Appendix contains further scores as a function of lead time.}
    \label{fig:overview}
\end{figure}

We start by investigating overall performance for 2m temperature predictions (Fig.~\ref{fig:overview}, Fig.~\ref{fig:app_scores_overview} in the Appendix), analysing different proper scores, that is, forecast verification metrics whose expected value is minimised when the forecast distribution matches the true distribution of the observations. We also analyse forecast calibration, i.e., whether the derived probabilities such as predictive intervals can be trusted, intuitively whether an event forecast with 30\% probability also happens in 30\% of similar forecast situations. All scores and metrics used are explained in Section \ref{sec:mat_methods}.

The probabilistic AIWP models outperform the raw IFS ensemble in all proper scores and at all lead times, and to a lesser degree also the EMOS-post-processed IFS. At 72h, WeatherNext2 improves the CRPS relative to the IFS by 25\% (90\% CI: 24--25\%), WeatherGen by 17\% (16--18\%) and EMOS by 14\% (13--14\%) (Fig.~\ref{fig:overview}a). Diebold--Mariano tests reject equal predictive performance for every model and lead time. The improvements are similar for the scaled CRPS \citep[][Fig.~\ref{fig:app_scores_overview}d]{bolin_local_2023}, which is less dominated by locations with large predictive spread, indicating substantial skill at low-variance locations as well. The gains are not restricted to small errors. The AIWP models reduce both the ensemble-mean RMSE and the magnitude and frequency of large errors (squared error $>4\,\mathrm{K}^2$; Fig.~\ref{fig:overview}c, Fig.~\ref{fig:app_scores_overview}f,h), with WeatherNext2 improving most. The improvements are consistent in terms of energy score (Fig.~\ref{fig:overview}b), indicating a better representation of the spatial structure of the forecast fields, and also show already in the first two forecast moments: in terms of the Dawid--Sebastiani score (Fig.~\ref{fig:app_scores_overview}e), WeatherNext2 outperforms EMOS, while WeatherGen performs on par with it, all three improving upon the raw IFS, meaning that they have better mean prediction and good forecast variance.

The AIWP and post-processed forecasts seem well calibrated, whilst IFS shows the characteristical underdispersion of NWP models for predictions of surface level variables \citep{vannitsem_2018_chapter_2}: spread-skill ratio of 0.65--0.7 at short lead times that increases to about 0.88 at 120h, see Fig.~\ref{fig:overview}d). Probability integral transform (PIT) histograms show that this underdispersion is mainly on the warm side: observations lie above all ensemble members far more often than below them (Fig.~\ref{fig:app_scores_overview}j). EMOS seems to outperform the AIWP models in terms of spread-skill ratio, however, PIT histograms suggest good calibration of all three models (Fig.~\ref{fig:app_scores_overview}j). Binned spread-skill diagrams (Fig.~\ref{fig:app_scores_overview}l) show that this calibration holds across ensemble spreads, meaning that forecasts are reliable also in uncertain situations, except for WeatherGen which has slight underdispersion when it predicts large spreads.

In terms of spatial patterns the CRPS improvements occur over most land areas and are largest over mountainous regions for all models (Fig.~\ref{fig:overview}e--g). WeatherNext2 improves almost uniformly, apart from a region in western Central Africa; WeatherGen has similar difficulties there and also performs worse than the IFS over parts of Central Africa and South America and towards the poles. EMOS shows weaker improvements throughout, with similar negative skill regions notably smaller gains over Eurasia and the eastern United States.

\subsection{AIWP models outperform the IFS for extremes, including records, and are calibrated in the tail}

\begin{figure}[t]
    \centering
    \includegraphics[width=1.\linewidth]{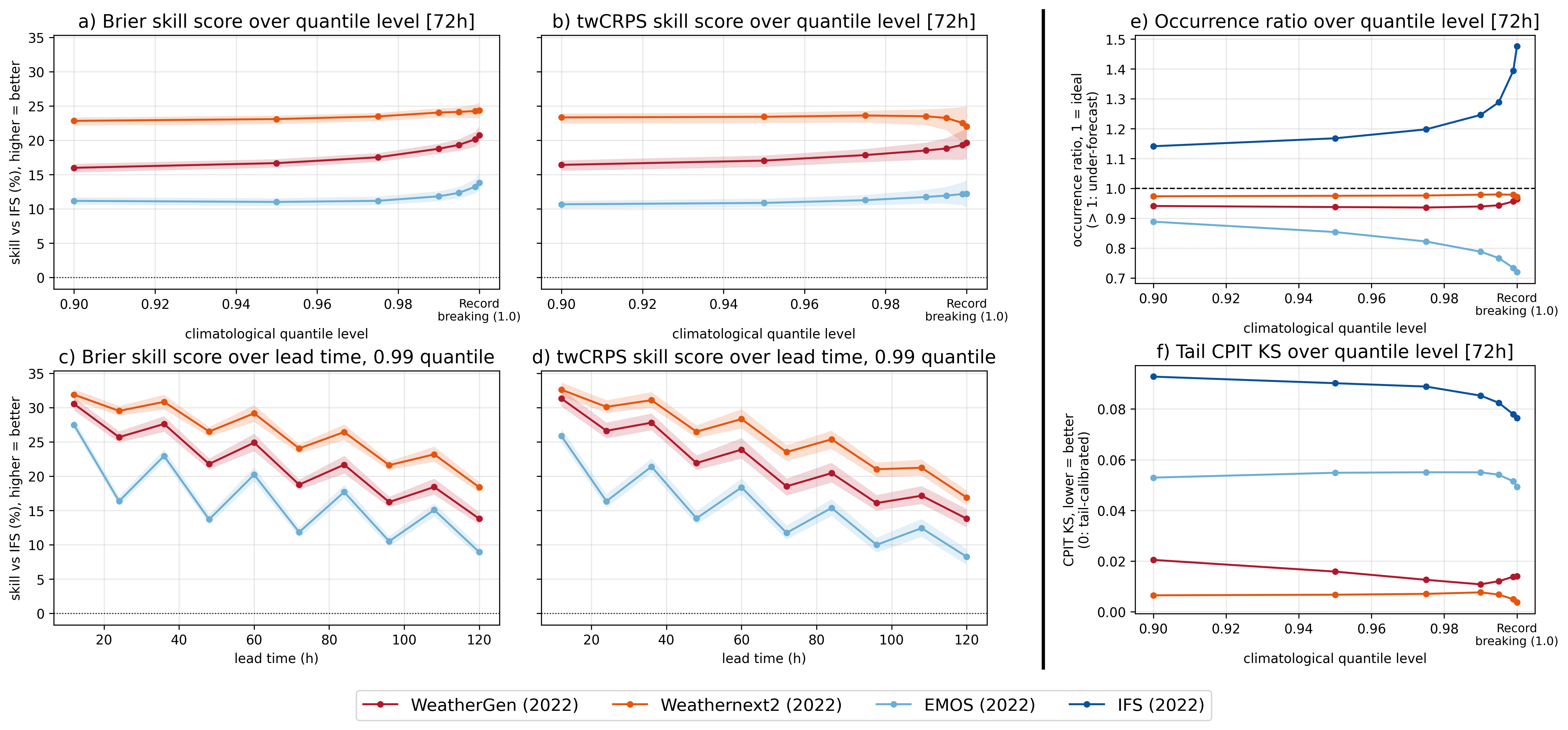}
    \caption{\textbf{Probabilistic scores for extreme events and tail calibration of AIWP and IFS ensemble, raw and post-processed.} Left: skill scores (\%) relative to the IFS, higher is better. a) Brier skill score and b) twCRPS skill score at 72h lead as a function of the climatological quantile level of the threshold; c) Brier skill score and d) twCRPS skill score for the 99th percentile as a function of lead time. Shading shows 90\% moving-block bootstrap intervals. Right: tail calibration at 72h lead as a function of the quantile level, e) occurrence ratio of observed to forecast exceedances (1: calibrated, $>1$: under-forecasting), f) Kolmogorov--Smirnov statistic of the conditional PIT values (0: tail calibrated). A quantile level of 1 corresponds to record-breaking events.}
    \label{fig:extremal_scores_cali}
\end{figure}

We now turn to analyse performance of the forecast ensembles at predicting of heat extremes, including record-breaking heat extremes. We analyse extremes in terms of exceedances of local thresholds as defined above. Note that exceedance of the 100th percentile threshold , thus, characterizes a record-breaking extreme.

We find that AIWP models outperform both IFS and post-processed IFS on predictions of temperature extremes and record-breaking extremes (Fig.~\ref{fig:extremal_scores_cali}a,b). AIWP models show better performance in terms of the Brier skill score (Fig.~\ref{fig:extremal_scores_cali} a,c), indicating a better prediction of the occurrence of heat extremes, including records: at 72h, WeatherNext2 achieves Brier skill scores over the IFS of 23\% at the 90th percentile and 24\% (23--26\%) for record-breaking extreme events. WeatherGen's skill increases from 16\% to 21\% (20--22\%), and EMOS's from 11\% to 14\% (13--15\%). This is consistent with results for the threshold-weighted CRPS (twCRPS, Fig.~\ref{fig:extremal_scores_cali} b,d), which evaluates occurrence and intensity predictions jointly, indicating that AIWP models are also better at predicting the intensity of events: skill over IFS is roughly constant across thresholds for WeatherNext2 (22--24\%) and EMOS (11--12\%) and increases for WeatherGen (16 to 20\%) as a function of the threshold. Diebold--Mariano tests reject the null hypothesis of equal predictive performance relative to the IFS in all 420 combinations of model, score, threshold and lead time (Benjamini--Hochberg correction at the 5\% level). The improvements in terms of Brier score and twCRPS are also consistent when considering individual seasons (not shown for sake of brevity).

Although WeatherGen and EMOS are close in terms of overall CRPS, both WeatherGen and WeatherNext2 clearly outperforms EMOS for extremes: Diebold-Mariano tests reject the null of equal predictive performance of AIWP models relative to EMOS in 279 out of 280 combinations (Benjamini--Hochberg correction at the 5\% level), the only exception being the Brier score for WeatherGen at the 90th percentile for 12h lead time. Post-processing thus improves the IFS by a roughly threshold-independent amount, but does not close the gap to the AIWP models. 

Skill decreases with lead time (Fig.~\ref{fig:extremal_scores_cali}c,d), from about 30\% at 12h to 14--18\% at 120h for the AIWP models, with a pronounced diurnal oscillation. Because all forecasts are initialised at 00UTC, lead times of $24i$h verify at night over most land areas (63.5\% percent of global land area) and lead times of $24i+12$h during the day. Skill relative to the IFS is consistently higher at the latter, i.e.\ during the day. The challenges of NWP forecasting of night-time stable boundary layer processes are well documented \citep[e.g.][]{holtslag_stable_2013}; the AIWP models carry a weaker form of the same diurnal signal, which they may inherit from the (re)analysis they are trained on. The spatial pattern of improvement in the Brier score at the 99.5th percentile (Fig.~\ref{fig:app_bs_maps}) is largest over Central Africa, where many exceedances occurred in 2022 and the IFS Brier score is high. This is notably close to where WeatherGen and EMOS perform worse in CRPS. Further improvements occur over mountainous and coastal regions, though not always where the CRPS skill is highest. These patterns are qualitatively similar for thresholds between the 90th percentile and records. Individual events, however, shape the results at individual gridpoints substantially; therefore the maps should be interpreted with care.

Predictions should also be calibrated, i.e., users should be able to trust the derived probabilities. This is particularly important for extremes, where high-stakes decisions are taken on the basis of forecast probabilities. Good performance in terms of scores does not guarantee this \citep{wilks_enforcing_2018, wessel_enforcing_2026}. We therefore assess tail calibration, that is the calibration for forecasts of heat extremes, of the different forecasting models (Fig.~\ref{fig:extremal_scores_cali}e,f). We find that the AIWP models are also well calibrated: The occurrence ratio, which measures over- or underprediction, of WeatherNext2 stays close to one for all thresholds, including records ($\approx0.97$), and WeatherGen's is only slightly below one ($0.94$--$0.96$). Both have very small conditional-PIT KS statistics ($<0.02$), so the forecast intensities of extremes are also reliable. This is notable for WeatherNext2, as it is CRPS-trained and this has been suggested to favour overly confident forecasts \citep{buchweitz_asymmetric_2025}. The raw IFS underforecasts the occurrence of extremes, increasingly so for higher thresholds (occurrence ratio 1.14 at the 90th percentile, rising to 1.48 for records), which is consistent with its overall underdispersion. EMOS, conversely, overforecasts exceedances (occurrence ratio decreasing from 0.89 to 0.72 for records), suggesting its Gaussian predictive distribution is too heavy-tailed for record-breaking temperature events. Both the IFS's underforecasting and, more weakly, the AIWP models' occurrence ratios oscillate diurnally with lead time, with the IFS much better calibrated during the day (Fig.~\ref{fig:app_tail_calibration_records}c).

\subsection{AIWP models miss fewer record-breaking extremes and predict their intensity better than the IFS}

\begin{figure}[t]
    \centering
    \includegraphics[width=\linewidth]{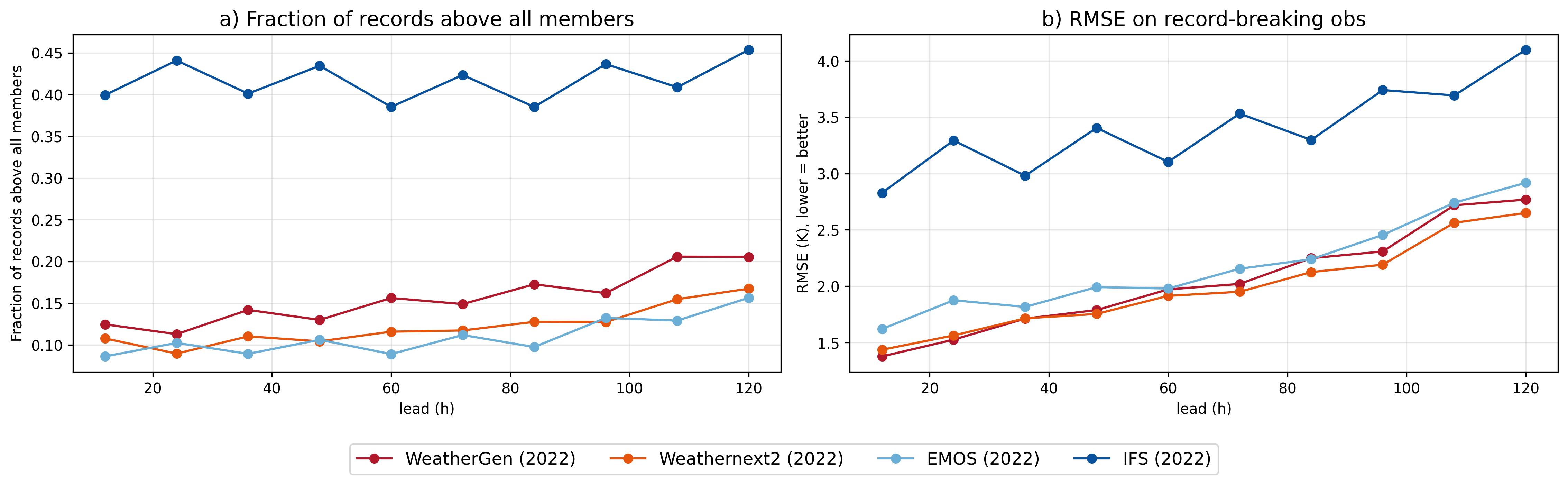}
    \caption{\textbf{Forecasts of record-breaking events, i.e.\ observations exceeding the 1979--2017 ERA5 maximum for the respective grid cell, month and hour.} a) Fraction of records lying above all ensemble members and b) RMSE of the ensemble mean on record-breaking observations, both as a function of lead time.}
    \label{fig:records}
\end{figure}

In the previous section, we found better performance of AIWP forecasts on record-breaking extremes in terms of occurrence and intensity. To further investigate forecast performance on records, we assess the extremes missed by the entire forecast ensemble (fraction of records above all members, Fig. \ref{fig:records}a), as well as the RMSE on record-breaking events only (Fig. \ref{fig:records}b).

We find that of all observed records, 39--45\% lie above every IFS ensemble member, compared with 9--17\% for WeatherNext2, 11--21\% for WeatherGen and 9--16\% for EMOS. A forecaster relying on the raw physical ensemble would therefore miss record-breaking extremes entirely at a much higher rate than forecasters relying on the AIWP ensembles. The ensemble-mean RMSE on record-breaking observations, which is not a proper measure but is commonly reported \citep{zhang_physics-based_2026}, gives the same ranking. The AIWP models and EMOS perform similarly (1.4--2.9\,K across lead times) and clearly better than the IFS (2.8--4.1\,K). The intensity of record-breaking heat extremes is therefore forecast more accurately by the AIWP ensembles. Notably, EMOS closes the gap to the AIWP forecasts in terms of these metrics, which is not the case in terms of proper tail-focused scores and tail-calibration.

We also evaluate the HRES forecast. On records, HRES lies above the IFS ensemble mean in more than 80\% of cases, often in the upper quintile of the ensemble and occasionally above its range. It therefore attains a much lower RMSE on records than the ensemble mean. This reflects the fact that a single physical integration does not shrink towards the conditional mean, rather than a superior forecast, as has been concluded in some studies \citep[e.g.][]{zhang_physics-based_2026}. Conditioning on observed records favours such a forecast, and it cannot be meaningfully compared to the output of an MSE-trained deterministic model, which estimates a different forecast target. Thus, taken together, these results suggest that at least some of the deficits reported for deterministic AIWP models reflect a choice of comparison, rather than a limitation of AIWP models. In terms of probabilistic scores the AIWP models outperforms and capture more record-breaking events within their ensemble range.

\subsection{The AIWP advantage holds across regions and in the July 2022 European heatwave}

\begin{figure}[t]
    \centering
    \includegraphics[width=\linewidth]{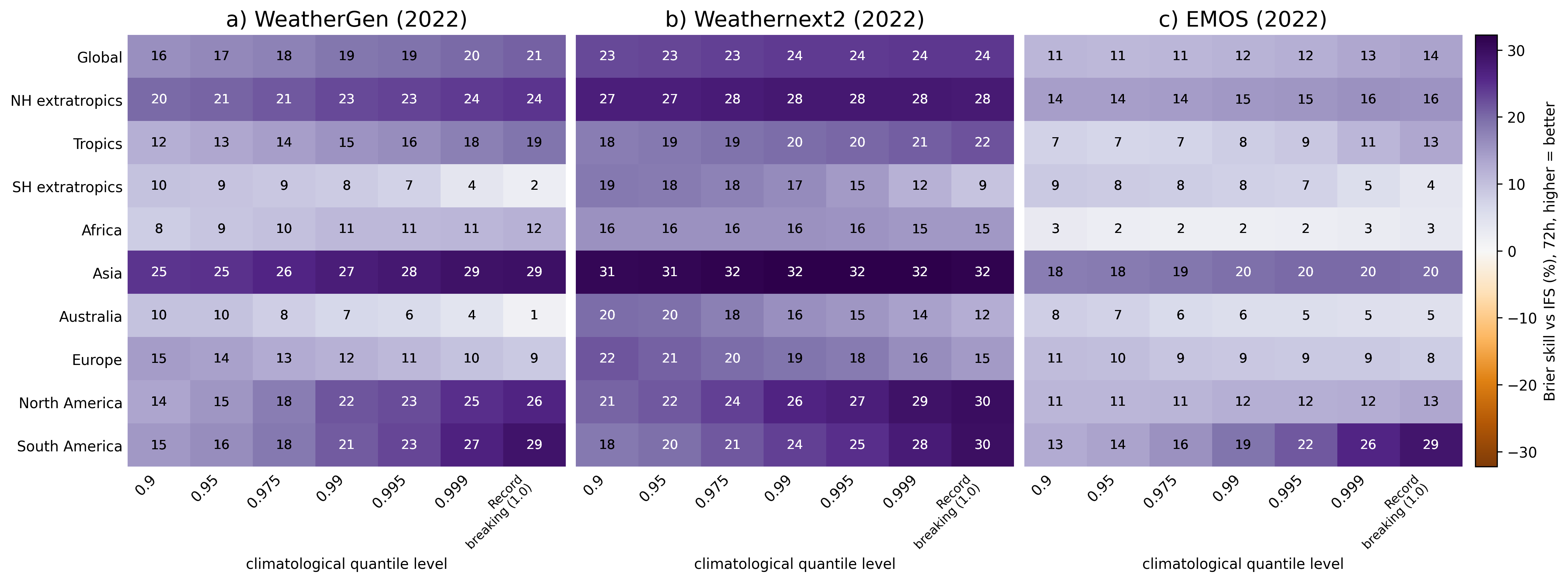}
    \caption{\textbf{Regional skill of AIWP models and EMOS over the raw IFS for record-breaking heat extremes}. Brier skill score (\%) at 72h lead time relative to the IFS for a) WeatherGen, b) WeatherNext2 and c) EMOS, for different regions (rows) and climatological quantile levels of the threshold (columns). A quantile level of 1 corresponds to record-breaking events. Positive values indicate an improvement over the IFS.}
    \label{fig:heatmap_bs}
\end{figure}

\begin{figure}[t]
    \centering
    \includegraphics[width=\linewidth]{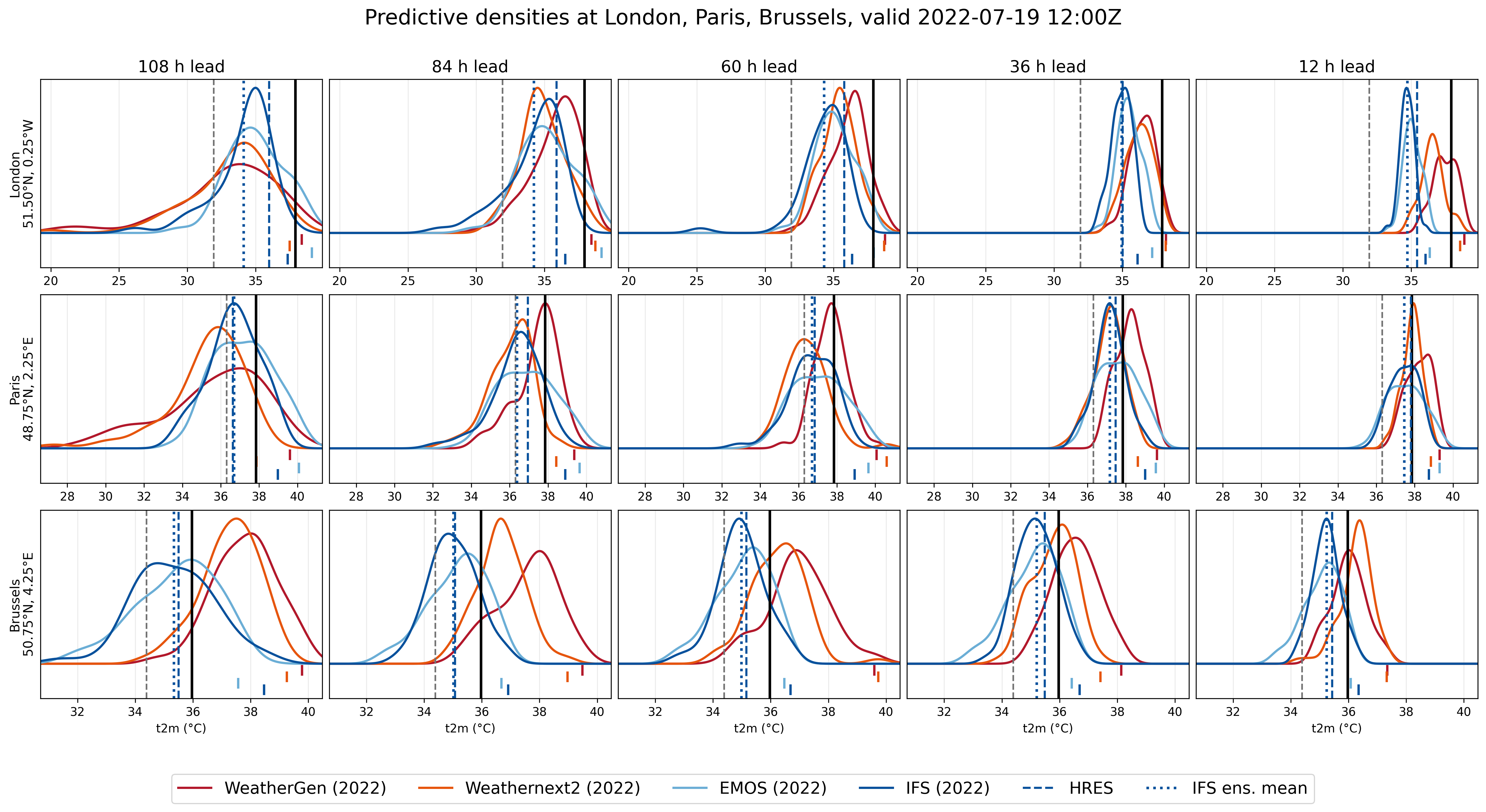}
    \caption{\textbf{Case study of the July 2022 heatwave in London, Paris and Brussels.} Predictive densities valid 19 July 2022 12UTC for London (top), Paris (middle) and Brussels (bottom), for lead times from 108h (left) to 12h (right). Densities are kernel density estimates of the ensemble members. The verifying HRES-fc0 observation is shown as a solid black line and the previous (1979--2017) local record as a thin grey dashed line. The HRES forecast is shown as a dashed brown line and the IFS ensemble mean as a dotted brown line. The highest ensemble member of each model is marked by a short tick in the model's colour.}
    \label{fig:densities}
\end{figure}

We have so far assessed globally aggregate metrics, which could mask relevant regional performance differences. We therefore conduct a regional analysis to complement the previous results, assessing performance on both continental and hemispheric scales (Fig.~\ref{fig:heatmap_bs}).

In terms of forecasting the occurrence of heat extremes, assessed using the Brier skill score, AIWP models outperform the raw IFS across all regions and thresholds considered and also hold a large edge over EMOS. However, there are considerable regional variations in this improvement: Skill is largest over Asia (WeatherNext2: 31--32\%) and, for the highest thresholds, over North and South America (up to 30\%). It is smallest over Africa and the Southern Hemisphere extratropics. In Australia and the SH extratropics, WeatherGen's skill decreases towards zero for records (1--2\%), whereas WeatherNext2 retains 9--12\%. EMOS skill is markedly smaller and over Africa it barely improves on the IFS (2--3\%), consistent with the negative CRPS skill there (Fig.~\ref{fig:overview}g). Over South America, EMOS skill increases strongly with the threshold and reaches the level of the AIWP models for records. These findings hold for a range of other scores such as miss rates and the twCRPS which are omitted here for sake of brevity. These improvements on a regional level are also consistent when disaggregating by lead time (Fig.~\ref{fig:app_skill_lead_region}), where AIWP models outperform IFS for record-breaking event predictions in all regions in terms of Brier score. They are also consistent when considering individual seasons in individual regions (not shown for sake of brevity).

In order to gain further insight into the predictions produced by IFS, EMOS and the AIWP models, we consider one case study: the July 2022 European heatwave and forecasts valid 19 July 12UTC, at London, Paris and Brussels. All three locations exceeded their 1979--2017 record. Figure~\ref{fig:densities} shows predictive densities of IFS and AIWP models. In London, all models issue similar forecasts at 108h lead, but as lead time decreases the IFS ensemble concentrates below the observation: at 36h and 12h the observed temperature lies above all IFS members. The AIWP models, particularly WeatherGen, keep substantial probability mass near the observed value. In Paris, all models capture the event, but the AIWP models, particularly WeatherGen, converge on the observation earlier, and WeatherNext2 produces markedly sharper densities at short lead times. In Brussels, the AIWP models initially overestimate the temperature but correct this at shorter lead times, whereas the IFS and EMOS underestimate the event at all lead times shown.

\subsection{Model rankings for extremes depend on the verification dataset}\label{sec:verif_data}

\begin{figure}[!t]
    \centering
    \includegraphics[width=\linewidth]{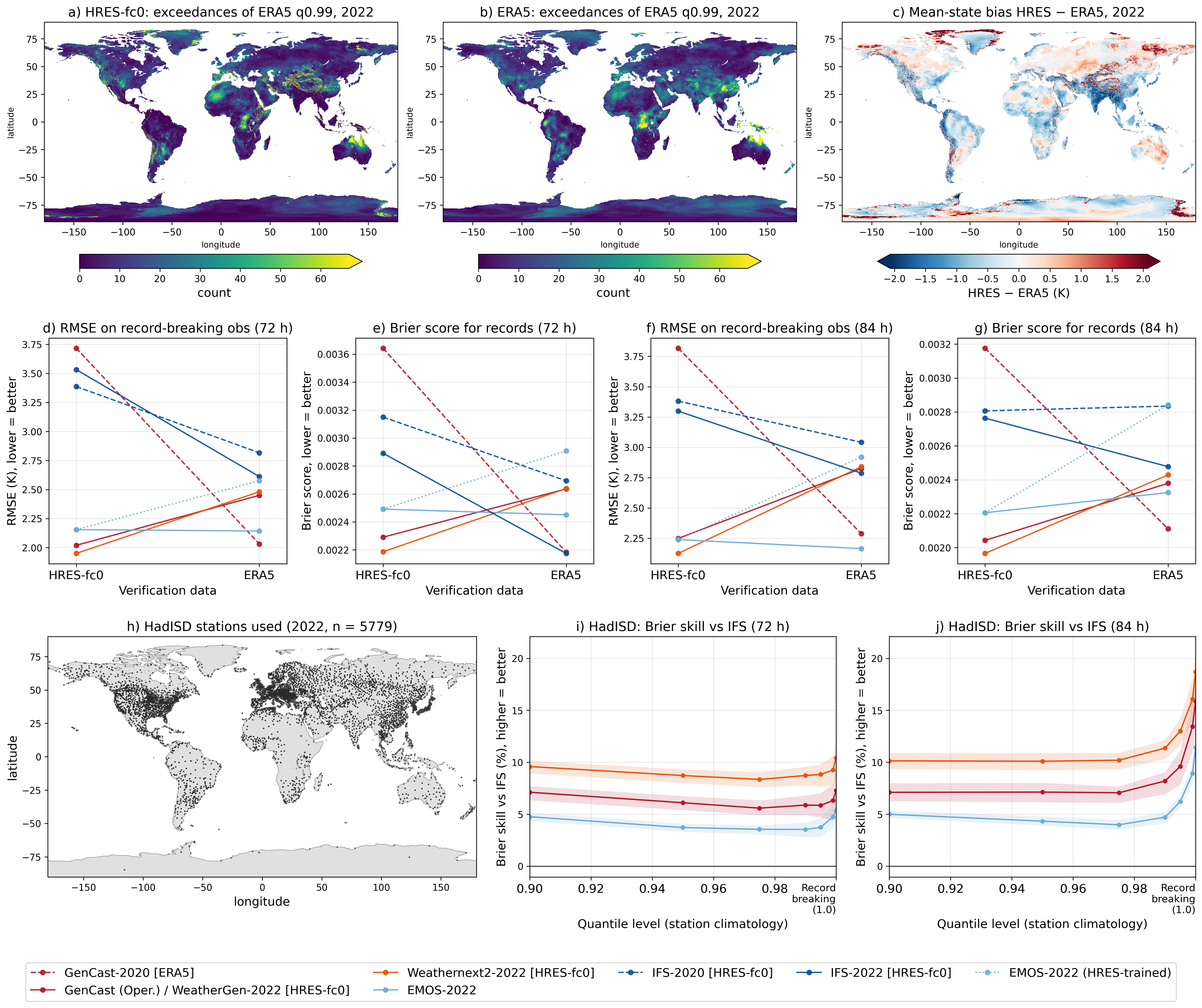}
    \caption{\textbf{Evaluation of the sensitivity of model rankings of AIWP models and IFS to the choice of verification dataset.} a,b) Number of exceedances of the ERA5 99th percentile (1979--2017, per grid cell, month and hour) in 2022 in a) HRES-fc0 and b) ERA5. c) Mean difference between HRES-fc0 and ERA5 in 2022. d--g) Ensemble-mean RMSE on record-breaking observations (d,f) and Brier score for record-breaking heat extremes (e,g) at 72\,h (valid 00\,UTC; d,e) and 84\,h (valid 12\,UTC; f,g), verified against HRES-fc0 (left) and ERA5 (right). h--j) Verification against HadISD station observations. h) Stations used in 2022 ($n=5778$). i,j) Brier skill relative to the IFS of the same year at 72\,h (i) and 84\,h (j) as a function of the quantile level of each station's own 1979--2017 climatology (level 1: record). Forecasts are taken from the nearest land grid cell. Shading shows 90\% moving-block bootstrap intervals.  EMOS-2022 is trained on the respective verification dataset (solid) or on HRES-fc0 for both (dotted). In h--j, EMOS-2022 is the HRES-fc0-trained version.}
    \label{fig:hres_vs_era}
\end{figure}

In this final section, we examine the sensitivity of our results to the choice of verification dataset. In AIWP verification, models are commonly verified against the dataset they were trained on or initialised from (ERA5 for ERA5-trained models, HRES-fc0 for the IFS and operational models), to avoid penalising them for systematic differences between analyses and to ensure a fair comparison against the IFS, which performs worse when scored on ERA5 \citep[e.g.][]{lam_graphcast_2022, rasp_weatherbench_2024}. So far, we have followed this convention and verified operational IFS and AIWP forecasts against HRES-fc0. However, when all models share the same initial conditions, the choice of verification dataset can still affect the scores. In this section, we therefore compare our results with verification against ERA5, which is widely used in other verification studies \citep[e.g.][]{pasche_validating_2025, zhang_physics-based_2026, dacre_northern_2026}, and against station observations.

Comparing ERA5 and HRES-fc0, we find that whether an event is classified as exceeding a threshold depends strongly on the dataset: HRES-fc0 and ERA5 frequently disagree even though both are compared with the same ERA5-derived thresholds. At the 90th percentile, 65\% of ERA5 exceedances are also exceedances in HRES-fc0, dropping to 51\% for records; conversely, the share of HRES-fc0 exceedances also found in ERA5 drops from 75\% to 40\%. In 2022, ERA5 contains 622\,799 record-breaking cases and HRES-fc0 795\,818, of which only 317\,268 coincide, although many are on similar days and nearby grid cells. The spatial distribution also differs (Fig.~\ref{fig:hres_vs_era}a,b). HRES-fc0 has many more exceedances over mountainous regions and along coastlines, a more spotty pattern, more events over Central Asia, and the distribution of exceedances within Africa differs between the two datasets. These differences partly reflect mean-state differences (Fig.~\ref{fig:hres_vs_era}c): HRES-fc0 is cooler than ERA5 over much of the tropics and over South and East Asia, and warmer over parts of the Arctic and northern mid-latitudes. Furthermore, the two models differ substantially over mountaineous terrain, which can be explained by their different resolution (ERA5 9\,km, HRES-fc0 31\,km)

These differences have consequences for model rankings (Fig.~\ref{fig:hres_vs_era}d,e; Appendix~\ref{app:hres_era_fcst}). Verified against HRES-fc0, the operational AIWP models clearly outperform the IFS at all thresholds in terms of Brier score and twCRPS, and also have lower RMSE and miss rates on records. Verified against ERA5, however, performance differences are inconsistent across metrics. In ERA5, both AIWP models have performance that is much more similar to the IFS, with better RMSE on records, miss rates and Brier score at lead times valid at 12\,UTC (day for the majority of the land surface, e.g. 84h lead time), but they are outperformed by the IFS in terms of Brier score records at 00\,UTC (e.g. 72h lead time). We note that most of the AIWP deficit against ERA5 arises from records that occur only in HRES-fc0, to which they assign substantially higher probabilities than the IFS and which count as hits against HRES-fc0 but as false alarms against ERA5.

To understand this better, we also analysed the ERA5-trained non-operational (i.e. non-HRES-fc0 initialised and fine-tuned) GenCast in 2020 and compared it to the IFS in 2020. It shows the opposite behaviour, performing worst of all models against HRES-fc0, but clearly outperforms the IFS on ERA5 for all extremal scores, including scores for records (Fig.~\ref{fig:hres_vs_era}) and on overall scores (Fig. \ref{fig:app_overview_all_against_era5}). It seems that for both verification datasets, there is an AIWP model that outperforms the IFS. However, AIWP models inherit the biases of the dataset they are trained on, and which model outperforms the IFS therefore depends on the model's training data. In particular, we find that the operational AIWP models which are trained on HRES-fc0, lose most of their advantage over mountainous regions when verified against ERA5, where they clearly outperform the IFS when verified against HRES-fc0 (Fig.~\ref{fig:app_bs_maps_all_era}). As highlighted above, this points to a representativeness issue in which HRES-fc0 resolves orographic temperature structure that ERA5 smoothes out, and models tuned to HRES-fc0 reproduce this structure, which counts as error when verified against ERA5. Overall, we therefore believe that HRES-fc0 is a more reliable estimate as it is produced with a more recent version of the IFS (cycle 47r3 in 2022, compared with 41r2 for ERA5), assimilates more and newer observations, and runs at much higher horizontal resolution.

Nonetheless, to resolve the performance differences identified, we also assess skill relative to a station-dataset which neither of the AI models was directly trained on, even though it is assimilated in both ERA5 and HRES-fc0: the HADISD  dataset \citep{dunn_hadisd_2012}, with extremes defined relative to each station's own 1979--2017 climatology. Against 5\,778 stations, the operational AIWP models again outperform the IFS for extremes (Fig.~\ref{fig:hres_vs_era}h--j). WeatherNext2 improves the Brier score for records by 10\% at 72\,h and 19\% at 84\,h, and WeatherGen by 7\% and 16\%, with smaller but significant gains at all lower thresholds. Both also miss fewer station records than the IFS (50--52\% compared with 62--63\% of records lying above all ensemble members). Station verification thus supports the conclusions obtained with HRES-fc0 for the operational models. 

The ranking of model performance, including on extremes and record-breaking extremes, therefore depends in part on the verification data, an important consideration in particular when studying record-breaking extremes not assessed in previous studies. A possible way forward to account for observation errors are observation-error-corrected scores, introduced by \citet{ferro_measuring_2017}. Evaluating models using the observation-error-corrected Brier scores, which represents the disagreement between ERA5 and HRES-fc0 as a stochastic misclassification process and accounts for the observational uncertainty between ERA5 and HRES-fc0, shows the AIWP models outperforming the IFS for extremes under either dataset as truth (Appendix~\ref{sec:obserr}).

\section{Discussion}

We have evaluated probabilistic AI weather prediction (AIWP) models for heat extremes over global land areas and compared them with the raw and post-processed IFS ensemble. Consistent with previous literature, we find both WeatherGen and WeatherNext2 to outperform the raw IFS in terms of average performance. Moreover, we here show that also for extremes, including records, the AIWP models perform better than physics-based numerical weather prediction (NWP). The improvements are significant for every lead time, threshold and score considered. Both AIWP models also outperform the ensemble model output statistics (EMOS)-post-processed IFS, which is closer to operational practice, with the largest margins in the tail. The advantage is not limited to proper scores. The AIWP models are well calibrated in the tail in both the occurrence and the intensity of extremes, and they miss far fewer record events than the corresponding IFS forecasts. The improvements are also consistent across regions.

These results contrast with recent claims that AIWP models underestimate the occurrence and intensity of record-breaking extremes relative to physics-based models \citep[e.g.][]{zhang_physics-based_2026}. We argue that much of these reported shortcomings reflects the design of the comparison rather than a limitation of AIWP models. Deterministic AIWP models, trained with a mean squared error, estimate the conditional mean of the predictive distribution \citep{gneiting_making_2011} which has a muted representation of extremes. Comparing this to HRES, the control member of the ensemble data assimilation, conflates forecast target with method performance. Instead we encourage probabilistic verification in which we find that AIWP models outperform on extremes including record-breaking ones. Our results do not rule out limitations of AIWP models in extrapolating beyond the training distribution. They do, however, show that such limitations do not currently translate into worse probabilistic forecasts of record-breaking heat than those of the operational physics-based ensemble.

Our results also show that the choice of verification dataset matters considerably more for extremes than for average performance. Two commonly used "ground-truths", HRES-fc0 and ERA5, disagree on a large fraction of threshold exceedances and records, so the set of events a model is evaluated on depends on the dataset chosen as truth. Model rankings for extremes change accordingly. Models trained or fine-tuned on HRES-fc0 outperform the IFS when verified against HRES-fc0 but not always at the highest thresholds against ERA5. In contrast, ERA5-trained AIWP models such as GenCast-2020 clearly outperforms the IFS against ERA5. However, when verified against station data the operational HRES-fc0 finetuned AIWP models strongly outperform IFS again, including on record-breaking events. Further under a stochastic representation of the disagreement between the datasets the AIWP advantage holds for both. Nonetheless, this means that conclusions about individual models should be reported together with the verification data they rest on. It also raises the broader question to what extent forecasting systems for extremes should be robust to the choice of analysis used as truth.

Some limitations of the present study should be noted. First, it is a global analysis on a 0.25$^\circ$ grid verified against analyses, whereas impacts of extremes are usually local. We believe more extensive station-based verification would be useful, possibly comparing station-post-processed AIWP and NWP models \citep[see][for a station-based evaluation not focused on extremes]{bremnes_evaluation_2024}. Second, EMOS serves only as a proxy for operational post-processing chains which are much more complex in praxis and include other more complex statistical post-processing, neighbourhood blending, forecast combination with other models, as well as judgments by operational meteorologists. Adequate verification should ideally be end-to-end as the raw model forms only part of the chain. As forecasting centres begin to integrate AIWP models into their blends \citep[e.g.][]{trotta_post-processing_2025}, such evaluations will become possible in the near term. Third, the IFS ensemble is evaluated in its 2022 configuration. Its horizontal resolution has since been increased, and whilst the results apply only to the operational systems available for this period, we hope that they might generalise to other configurations and NWP/AIWP ensembles. Finally, we analyse a single year of operational forecasts and a single variable. Near-surface temperature is spatially and temporally smooth, and provides the large samples needed for statistical evaluation of the tails. For rarer high-impact events such as tropical cyclones or extreme precipitation, statistical evaluation might be limited by sample size, and it remains open whether AIWP models can forecast such events beyond the historical record. Further, results for individual regions are shaped strongly by the events recorded in 2022, analysis of AIWP performance on a longer dataset would be useful. 

Understanding why AIWP models perform well, or fail, for extremes thus requires further work. We believe that probabilistic evaluation with proper scores and calibration metrics can provide one line of evidence and that this paper can serve as guidance for researchers. However, many other approaches to understanding the behaviour of AIWP models at extremes are also useful. For example, experiments in which extreme events are removed from the training data and models retrained can show to what extent forecasts of extremes rely on similar events in training \citep{sun_can_2025}. Evaluations relative to location-wise thresholds, as here, could be complemented by analyses of spatial generalisation \citep{sun_predicting_2025}. Further, it may be worth leveraging large datasets of climate model ensemble or rare event simulations to see whether AIWP models can forecast non-observed realisations of internal variability \citep[e.g.][]{lancelin_ai-boosted_2025}.

\section{Materials and Methods}\label{sec:mat_methods}

\subsection{Data}

We rely on data from WeatherNext2 and the publicly available WeatherBench2 archive \citep{rasp_weatherbench_2024}, which provides forecasts and verification data on common grids. We evaluate surface temperature predictions from two probabilistic AI weather prediction models: operational GenCast / WeatherGen \citep{price_probabilistic_2025}, FGN / WeatherNext2 \citep{alet_skillful_2025} which we benchmark against raw (IFS) and post-processed IFS (IFS-EMOS). 

\paragraph{Forecasts and verification.}
We consider 2m temperature (t2m) forecasts from GenCast \citep{price_probabilistic_2025}, FGN \citep{alet_skillful_2025}, IFS and IFS-EMOS at 0.25° resolution in 2022. For all models we consider forecasts initialised at 00UTC and lead times of 12-120h in 12h steps. We restrict the evaluation to land, defined as grid cells with an ERA5 land--sea fraction exceeding 0.5, leading to 351848 of 1038240 cells. Ensemble sizes are 56 (GenCast), 64 (FGN), 50 (NeuralGCM) and 50 (IFS-ENS). The raw (and hence also post-processed) IFS, as well as GenCast and FGN are all in operational model and are are initialised from the operational ECMWF ensemble data assimilation and it's control member respectively. 

We verify AIWP and IFS forecasts against HRES-fc0, the 0th hour HRES forecast from the 4D-Var analysis, as is standard when comparing operational AIWP models and the IFS \citep[e.g.][]{lam_graphcast_2022, bi_accurate_2023, chen_fuxi_2023, rasp_weatherbench_2024, zhang_physics-based_2026}. HRES-fc0 provides the best estimate of surface conditions and is generated by the same model as the IFS, comparing against it instead of ERA5 therefore does not unduly favour the ERA5-trained AIWP models. In addition, however, we also evaluate against ERA5 and quantify differences in scores for extremes between ERA5 and HRES-fc0 (see Section \ref{sec:verif_data} and Appendix \ref{app:hres_era_fcst}). Finally, we also evaluate against the HADISD \citep{dunn_hadisd_2012} station dataset in Section \ref{sec:verif_data}. There we match each station with forecasts from the closest land grid cell if this cell is closer than 30km (removing some coastal and island stations). We also filter stations by completeness, using month-hour combinations if at least 400 valid observations from at least 20 different years are available.

Finally, in the Appendix we include additional results for NeuralGCM \citep{kochkov_neural_2024}, which is available only at 1.5° and carries no 2m temperature field; we therefore evaluate it on upper-air temperature at the 850hPA level and compare it against IFS forecasts of this variable. NeuralGCM is not available in operational mode and is initialised from ERA5, thus we compare it against this dataset. Further, we include results for non-operational GenCast, also initialised and compared against ERA5. Finally, we also quantify uncertainty in the verification product by evaluating all models also against ERA5 in addition to HRES-fc0 in the Appendix.

\paragraph{Statistical post-processing.}
Statistical post-processing is routinely applied in meteorological centers as raw ensemble forecasts are frequently biased and underdispersed, particularly for surface variables \citep{vannitsem_2018_chapter_2}. In addition AI models can be much more flexible with respect to surface level variables, whilst the IFS is constrained by physical parametrizations. We thus also consider the IFS-ENS post-processed with ensemble model output statistics \citep[EMOS;][]{gneiting_calibrated_2005}, a popular parametric post-processing technique that is employed operationally by several meteorological services \citep{roberts_improver_2023}. We fit a Gaussian EMOS model independently for each grid cell and lead time on IFS-ENS forecasts from 2018 and 2019, and apply it to 2022. Writing $m$ and $s^{2}$ for the ensemble mean and variance and $d$ for the day of year, the predictive distribution is
\begin{align}
  Y \mid m, s^{2} \;&\sim\; \mathcal{N}\!\left(\mu, \sigma^{2}\right), \\
    \mu &= a + b\,m + c_{1}\sin\!\left(\tfrac{2\pi d}{365.25}\right)
                    + c_{2}\cos\!\left(\tfrac{2\pi d}{365.25}\right), \\
    \sigma^{2} &= \gamma + \delta\,s^{2}, \qquad \gamma,\delta \ge 0 ,
\end{align}
where the annual harmonics absorb residual seasonal bias that the ensemble mean does not remove. The location coefficients are estimated by least squares and the two variance coefficients by regressing squared residuals on $s^{2}$. We note that maximum likelihood or CRPS-based fitting is more common \citep{gebetsberger_estimation_2018, wessel_lead-time-continuous_2024}, however, we choose the moment-based fitting as it is much faster. After fitting we build ensembles with M = 51 members by sampling from the normal predictive distribution. This is so that the post-processed forecast can be scored with the same ensemble-based estimators as the raw models.

\subsection{Verification methods}
We evaluate using different proper scores and calibration metrics, outlined in the following. According to \citet{gneiting_probabilistic_2007} probabilistic forecasts should be maximally sharp, subject to calibration. Sharpness is corresponds to the informativeness of a forecast, whilst calibration captures the statistical compatibility of forecast and observations and is a necessary condition to be able to trust the probabilities and predictive intervals derived from a probabilistic forecast. The condition of sharpness subject to calibration holds for the full distribution, but also for forecasts of extremes. We assess both calibration and sharpness jointly by evaluating the forecasts using proper scoring rules: these are probabilistic scores that are minimal at the true distribution \citep{gneiting_strictly_2007, waghmare_proper_2025}. We additionally investigate calibration and calibration of extreme event forecasts, as good performance in terms of scores might be achieved with gains in sharpness at the expense of calibration \citep{wilks_enforcing_2018, wessel_enforcing_2026}. 

\subsubsection{Definition of extreme events and threshold selection}
For the purpose of this paper we evaluate heat extremes as thresholds exceedances of location-wise thresholds corresponding to large quantiles of the historical training distribution. We define thresholds as gridcell-wise quantiles of the ERA5 distribution of 1979-2017, the period usually utilized for AIWP training. We define individual thresholds per month and relative to the forecasted hour. Concretely, for month $m$, grid-cell $i$, hour of the day $h$, probability level $p$, we define thresholds:
$$t_{i, m, h} = Q_{i,m, h}(p),$$
where $Q_{i,m,h}$ is the quantile function of the ERA5 distribution at this location and month. For analysing extremes we consider thresholds in the 90, 95, 97.5, 99 and 99.9th percentile of the historical distribution. We define \textit{record breaking heat extremes} as exceeding the 100th quantile of the historical distribution. This largely follows \citet{zhang_physics-based_2026}, but we note that our protocol differs slightly as they define extremes at the maximum value of ERA5 at 00, 06, 12, 18\,UTC at each gridcell. Our hour-matched definition instead allows to assess night-time heat extremes as well. For the  station-based analysis in Section~\ref{sec:verif_data} we define thresholds analogously, by considering quantiles of the 1979-2017 distribution for each station, month and forecasted hour. We note that by conditioning on such fixed thresholds this does not condition on observed outcomes whilst still looking at record breaking heat extremes, thereby circumventing the forecaster's dilemma \citep{lerch_forecasters_2017}.

\subsubsection{Proper scores}
We evaluate all forecasting models in terms of a number of proper scores. All location-wise scores $S$ are aggregated by computing global (or regional) averages with latitude weighting, accounting for grid cell size:
\begin{equation}\label{eq:agg}
    \bar S = \frac{\sum_{\tau=1}^{T} \sum_{i \in \mathcal{L}} w_i \, S_{i\tau}}{T \sum_{i \in \mathcal{L}} w_i},
    \qquad w_i = \cos(\phi_i),
\end{equation}
where $\phi_i$ is the latitude of cell $i$, so that $w_i$ is proportional to the cell's area. This aggregation ensures that $\bar{S}$ remains a (strictly) proper score.

\paragraph{Marginal scores.} As measure of average performance we consider the continuous ranked probability score \citep[CRPS; ][]{matheson_scoring_1976} for forecast CDF $F$ and observation $y$: 
\[
\mathrm{CRPS}(F,y) = \int_{-\infty}^\infty \left[F(z) - \one\{y \leq z\}\right]^2 dz = \E_{X\sim F} |X-y| -\frac{1}{2} \E_{X, X'\sim F} |X-X'|
\]
The ensembles have differing number of members: 50 for the ECMWF-IFS, 56 for WeatherGen, 64 for WeatherNext2, 51 for EMOS. Thus, we use the fair estimators throughout \citep{ferro_fair_2014}. Instead of scoring the forecast distribution spanned by the ensemble these give an estimate of the CRPS of the underlying distribution and are thus insensitive towards ensemble size. The CRPS can be sensitive towards the scale of the variable scored and might thus give larger importance towards locations with more uncertainty. Thus we also investigate the scaled CRPS, \citep{bolin_local_2023}:
\[
\mathrm{SCRPS}(F,y)
=
\frac{\E_{X \sim F} |X - y|}{\E_{X, X' \sim F} |X - X'|}
+
\frac{1}{2}
\log \left( \E_{X, X' \sim F} |X - X'| \right), 
\]
and verify whether model rankings agree between CRPS and SCRPS. The CRPS and SCRPS are strictly proper scoring rules. For the SCRPS we use the fair (unbiased) estimators of $\E|X-y|$ and $\E|X-X'|$, even if the full score retains a small finite-ensemble bias. We also evaluate the mean squared error of the ensemble mean, which is proper. Finally, we analyse the Dawid-Sebastiani score \citep[DSS;][]{dawid_coherent_1999}:
\[
\mathrm{DSS}(F,y)
=
\sum_{i=1}^D
\left(
\frac{(y_i - \mu_i)^2}{\sigma_i^2}
+
\log \sigma_i^2
\right),
\qquad
\mu_i = \E(\Xvec_i),\;
\sigma_i^2 = \mathrm{Var}(\Xvec_i).
\]
This is a proper score, analysing a forecast only in terms of the first two moments and quantifies the accuracy of those. We note that we are not aware of a fair version of the DSS, except for the specific case of a Gaussian forecast distribution \citep{siegert_ensembleadjusted_2019}, thus we use the standard version here. 

\paragraph{Tail-focused scores.} We evaluate extremes in terms of tail focused scores. For quantifying the performance in forecasting the occurrence of extremes, we consider the Brier score \citep[BS;][]{brier_verification_1950}, of which the CRPS is the continuous integration, given as:
\[
\text{BS}(F,y) = \left[F(t) - \one\{y \leq t\}\right]^2,
\]
we utilize again the fair estimator \citep{ferro_effect_2008}. Forecasts of the intensity of extremes should be evaluated jointly together with the occurrence \citep{lerch_forecasters_2017}. To do so we consider the threshold-weighted CRPS \citep[twCRPS;][]{gneiting_comparing_2011}:
\[
\text{twCRPS}_w(F,y) = \int_{-\infty}^\infty w(z) \left[F(z) - \one\{y \leq z\}\right]^2 dz = \E_{X\sim F} |v(X)-v(y)| -\frac{1}{2} \E_{X, X'\sim F} |v(X)-v(X')|.
\]
We use threshold weighting $w(z) = \one\{z\geq t\}$ for which the so-called chaining function becomes $v(z) = \max(t, z)$. This corresponds to a censoring of both forecasts and observations at $t$. The twCRPS focuses inference on events over a threshold $t$, whilst remaining proper and evaluates both the occurrence and intensity of such events. Similarly we use the fair twCRPS estimator.

\paragraph{Multivariate scores.} We also evaluate spatial structure of the forecasts in terms of Energy score, the multivariate generalization of the CRPS. The energy score is given as:
\[
\text{ES}(F,y)= \E_{X\sim F} \|X-y\| -\frac{1}{2} \E_{X, X'\sim F} \|X-X'\|.
\]
Classically an l2 norm $\|\cdot\|$ would be used, however, this is skewed by unequal grid cell size, thus we use the weighted norm $\|\mathbf{x}\|_w = \big(\sum_i \tilde w_i x_i^2\big)^{1/2}$ with normalised latitude weights $\tilde w_i = w_i / \sum_j w_j$, see \eqref{eq:agg}. This remains a strictly proper score as the kernel is strongly negative definite \citep{waghmare_proper_2025}.

\paragraph{Diebold--Mariano test.} We use the Diebold--Mariano test, to test whether score differences between AIWP models and the IFS are significant \citep{diebold_comparing_1995, vannitsem_2018_chapter_6}. Given model $A$ and baseline $B$ one can calculate the score difference $d_\tau = \bar{S}^A_\tau - \bar{S}^B_\tau$. Then, under the null hypothesis of equal predictive performance $\E[d_\tau] = 0$,  the statistic
\[
\text{DM} = \frac{\bar{d}}{\sqrt{\widehat{\sigma}^2_{\text{HAC}} / T}},
\]
is asymptotically standard normal. Here, we use the Bartlett kernel long-run variance:
\[
\qquad
\widehat{\sigma}^2_{\text{HAC}} = \widehat{\gamma}_0
+ 2\sum_{k=1}^{L}\left(1 - \frac{k}{L+1}\right)\widehat{\gamma}_k ,
\]
with $\widehat{\gamma}_k$ the lag-$k$ autocovariance of $d_\tau$ and $T$ the number of paired dates. This accounts for the serial dependence of the loss differential; we set the truncation lag to $L = h - 1 + 4$ days, where $h$ is the lead time in days, combining the usual $h-1$ correction for overlapping forecast windows with four additional days to accommodate synoptic persistence. Negative values of DM indicate that $A$ attains the lower, i.e.\ better, score. We test each model against its baseline for the twCRPS and Brier score at each threshold level and lead time. As this constitutes several hundred tests, we control the false discovery rate at the $5\%$ level using the Benjamini-Hochberg \citep{benjamini_controlling_1995} procedure and report adjusted $p$-values throughout.

\paragraph{Confidence intervals.} Interval estimates for mean scores and skill scores are obtained by a moving-block bootstrap over initialisation dates, which resamples blocks of consecutive days and thereby preserves the short-range temporal dependence that an independent bootstrap would destroy. We use a block length of five days, matching the synoptic timescale, and $B = 2000$ resamples, and report percentile intervals at the $90\%$ level. The skill score
\[
\text{Skill-score} = 1 - \frac{\sum_\tau \sum_i w_i S(F^A_{i,\tau}, y_{i,\tau})}
                     {\sum_\tau \sum_i w_i S(F^B_{i,\tau}, y_{i,\tau})}
\]
is computed from paired resamples $\tau$. We verified that our conclusions are insensitive to the choice of block length by repeating the procedure with block lengths of one and ten days.

\subsubsection{Calibration and tail calibration}

Calibration is a necessary requirement to trust probabilistic forecasts. Different notions of calibration have been proposed in the literature \citep{gneiting_probabilistic_2007, gneiting_regression_2023}. Most frequently assesed is probabilistic calibration, defined as uniformity of the PIT transform: 
\[
F(Y) \sim \text{Unif}[0,1].
\]
PIT, or the discrete alternative, rank histograms should be flat under probabilistic calibration which subsequently implies correct coverage of predictive intervals i.e. that a 90\% predictive interval covers the observations 90\% of the time \citep{vannitsem_2018_chapter_6}. We also investigate spread-skill ratios: the ensemble standard deviation $s$, divided by the root mean squared error of the ensemble mean:
\[
\mathrm{SS} = \frac{s}{\mathrm{RMSE}(F,y)}.
\]
This should be equal to one for a calibrated forecast. We use the fair mean squared error estimator \citep{fortin_why_2014}. In addition to computing this across all forecast cases we analyse the calibration for different levels of spread, by binning across values of $s$.

Probabilistic and spread-skill calibration, is a fairly weak notion of calibration as it is unconditional. In fact, it has been observed that forecasts might have good calibration overall, but bad calibration for extremes \citep{allen_tail_2025, wessel_enforcing_2026, wessel_towards_2026}. For example, forecasts might be calibrated overall, but might be overconfident when predicting the intensity high impact events or over/underestimate their occurrence. This means that the coverage of predictive intervals could be good when aggregated, but the same intervals might not be trustworthy for extremes. To address this \citet{allen_tail_2025} introduce a notion of tail calibration to assess the reliability of extreme event forecast. Tail calibration is defined as two conditions:
\begin{align}
    \quad \Q (Y\geq t) = \E [1-F(t)] \label{eq:occ}\\
    F_t(Y) \sim \text{Unif}[0,1] \text{ for } Y > t,  \label{eq:cpit}
\end{align}
where $F_t$ is the excess distribution:
\[
F_t(y) := \frac{F(y) - F(t)}{1-F(t)}.
\]
The first condition \eqref{eq:occ} assesses the reliability of occurrence forecasts of extremes: under tail calibration extremes should occur at the forecasted frequency. The second condition \eqref{eq:cpit} concerns the reliability of intensity forecasts of the conditional excess distribution $F_t(\cdot)$: conditional PIT values $F_t(Y)$ should be uniform which ensures that prediction intervals from this distribution have the correct coverage. To assess this in practice we consider the occurrence ratio of realised over expected events as well as histograms of conditional PIT (CPIT) values which should be uniform. We use randomized CPIT values to account for the discrete nature of the ensemble and differing ensemble sizes, which particularly affect CPIT values in the tail \citep{vannitsem_2018_chapter_6}. We also quantify the distance of the samples of CPIT values $z_1, \dots, z_n$ from uniformity by considering an estimate of the Kolmogorov-Smirnov statistic:
\[
\hat{\mathrm{KS}} = \sup_{i = 1, \dots, n} | \hat{R}_z(i/n) - i/n|,
\]
where $\hat{R}_z$ corresponds to empirical CDF of CPIT values.

\section*{Acknowledgements}
J.W. and S.S. acknowledge funding provided by the Heinz Maier-Leibnitz Prize of the German Research Foundation, the climXtreme project funded by the German Federal Ministry of Education and Research (Phase 2, project PATTETA, grant number 01LP2323C) and the EU Horizon project ‘Artificial Intelligence for Enhanced Representation of Processes and Extremes in Earth System Models’ (AI4PEX; grant agreement 101137682). We thank Google for making openly available the WeatherBench2 and WeatherNext2 datasets. The authors used the generative AI models Claude Opus 4.8 and Claude Opus 5 in Claude Code to support with writing of research code, as well as for editing of the manuscript. All outputs have been carefully checked and code has been tested for correctness. We take full responsibility for the final content of this work.

\section*{Author Contributions}

JW, FS and SS conceptualised the study. JW conducted the formal analysis and wrote the initial draft. All authors provided feedback on methods and supported with review and editing.

\section*{Code and data availability}

The forecast and gridded verification data is available from the WeatherBench2 \citep{rasp_weatherbench_2024} and the WeatherNext2 datasets \citep{google_weathernext2_2026}. The HADISD dataset is available from \url{https://www.metoffice.gov.uk/hadobs/hadisd/} (last accessed 06.10.2026). The authors will make the code to reproduce all results openly available upon acceptance.

\clearpage
\appendix
\renewcommand{\thefigure}{A\arabic{figure}}
\setcounter{figure}{0}

\section{Further Results}
We show some further results to the main paper. In addition to considering the operational WeatherNext models, here we also evaluate all the probabilistic AIWP models on WeatherBench2 \citep{rasp_weatherbench_2024}. We evaluate GenCast, non-operational (ERA5-trained and initialized) for 2020 and NeuralGCM \citep{kochkov_neural_2024} which does not have a t2m field, thus we evaluate the t850hpA forecasts. We follow the usual praxis of evaluating ERA5-trained models against ERA5 and the operational models against HRES-fc0 \citep[e.g.][]{lam_graphcast_2022, bi_accurate_2023, chen_fuxi_2023, zhang_physics-based_2026}. To provide benchmarks for the 2020 models, we also analyse the IFS in 2020.

The take-aways are largely similar as in the main paper. NeuralGCM is better than corresponding IFS forecasts, except for early lead times, where it has bad calibration and is underspread, which might be due to it's noise injection schedule. ERA5-trained GenCast verified against ERA5 has broadly similar performance as GenCast (Oper.) and is still very similar as EMOS. For extremes, GenCast (ERA5) shows a large improvement relative to ERA5-verified IFS in terms of proper scores and miss rates. NeuralGCM does outperform the IFS baseline for most thresholds, but tests for equal predictive performance are not rejected for records.  GenCast (ERA5) has better tail calibration than IFS, but a slight overforecasting bias, even though it shows good CPIT calibration (intensity). Neural GCM improves over IFS baseline in terms of tail calibration and in particular it shows good occurrence calibration.

\begin{figure}[t]
    \centering
    \includegraphics[width=1.\linewidth]{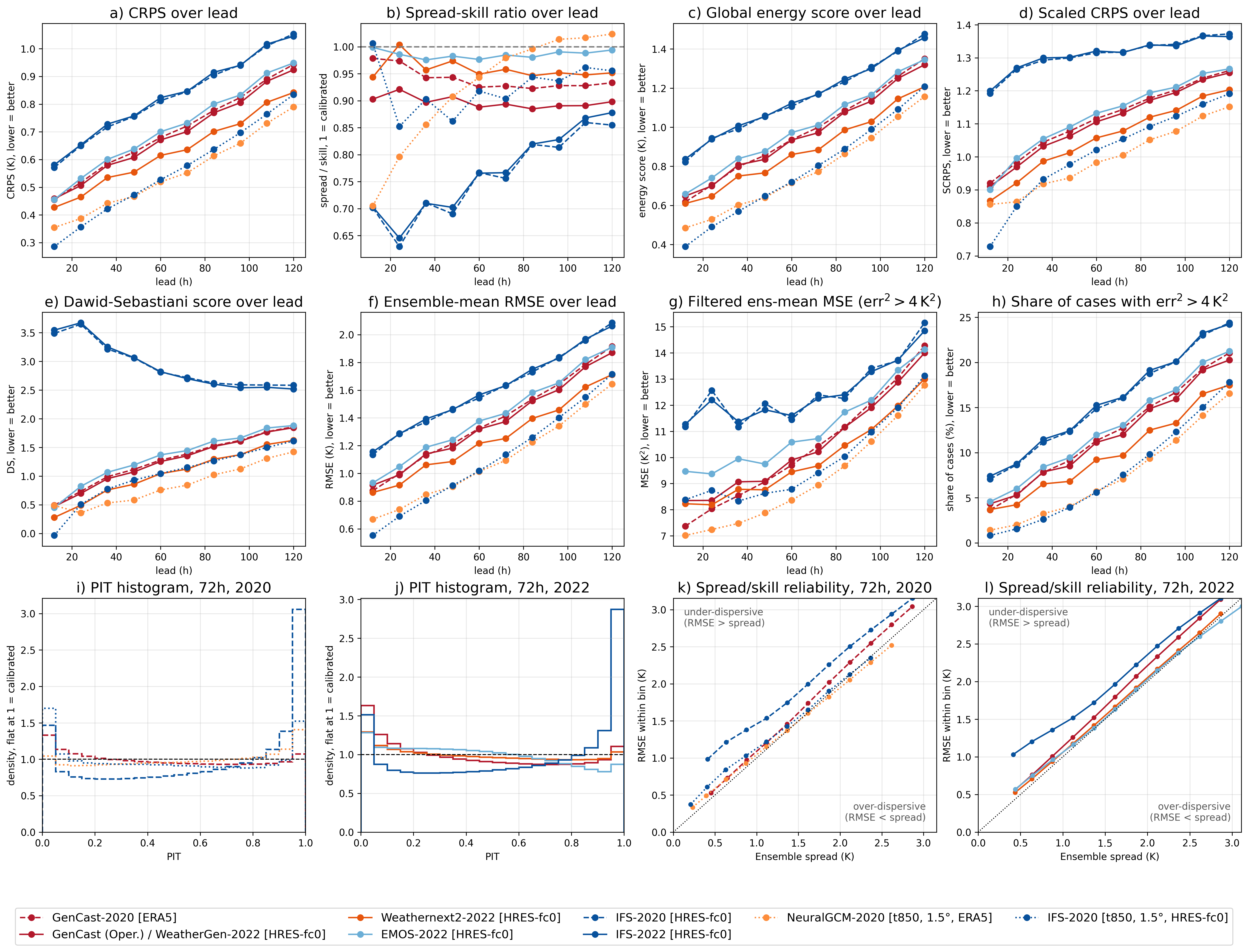}
    \caption{As Fig.~\ref{fig:overview}, additionally including the WeatherBench2 models GenCast-2020 (ERA5-trained and verified) and NeuralGCM-2020 (t850 at 1.5$^\circ$, verified against ERA5, with the IFS at the same level and resolution as reference). As a function of lead time: a) CRPS, b) spread-skill ratio, c) energy score, d) scaled CRPS, e) Dawid--Sebastiani score, f) ensemble-mean RMSE, g) ensemble-mean MSE restricted to squared errors $>4\,\mathrm{K}^2$ and h) fraction of cases with squared error $>4\,\mathrm{K}^2$. Calibration at 72h lead: PIT histograms for i) 2020 and j) 2022, and spread-skill reliability diagrams (RMSE within spread bins against ensemble spread) for k) 2020 and l) 2022. PIT histograms and spread skill ratios were split up for 2020 and 2022 for visibility.}
    \label{fig:app_scores_overview}
\end{figure}

\begin{figure}[t]
    \centering
    \includegraphics[width=1.\linewidth]{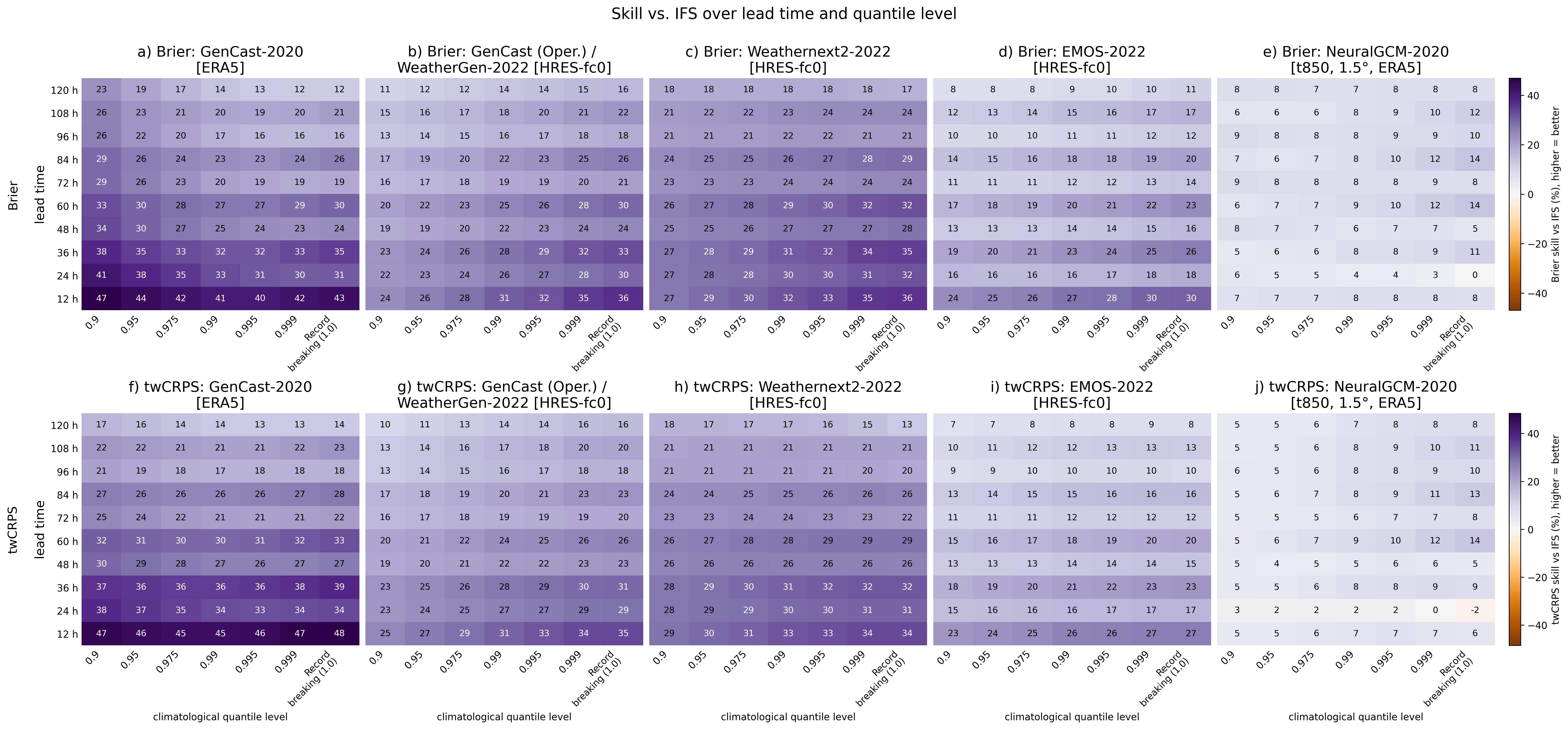}
    \caption{Skill (\%) relative to the IFS as a function of lead time (rows) and climatological quantile level (columns), for the Brier score (top) and twCRPS (bottom), for a,f) GenCast-2020, b,g) WeatherGen, c,h) WeatherNext2, d,i) EMOS and e,j) NeuralGCM-2020. Positive values indicate an improvement over the IFS.  Square brackets give the verification data. Note that the skill scores for the models verified against ERA5 are likely to be overoptimistic as a measure of performance as they are compared against ERA5-verified IFS.}
    \label{fig:app_skill_heatmap}
\end{figure}

\begin{figure}[t]
    \centering
    \includegraphics[width=1.\linewidth]{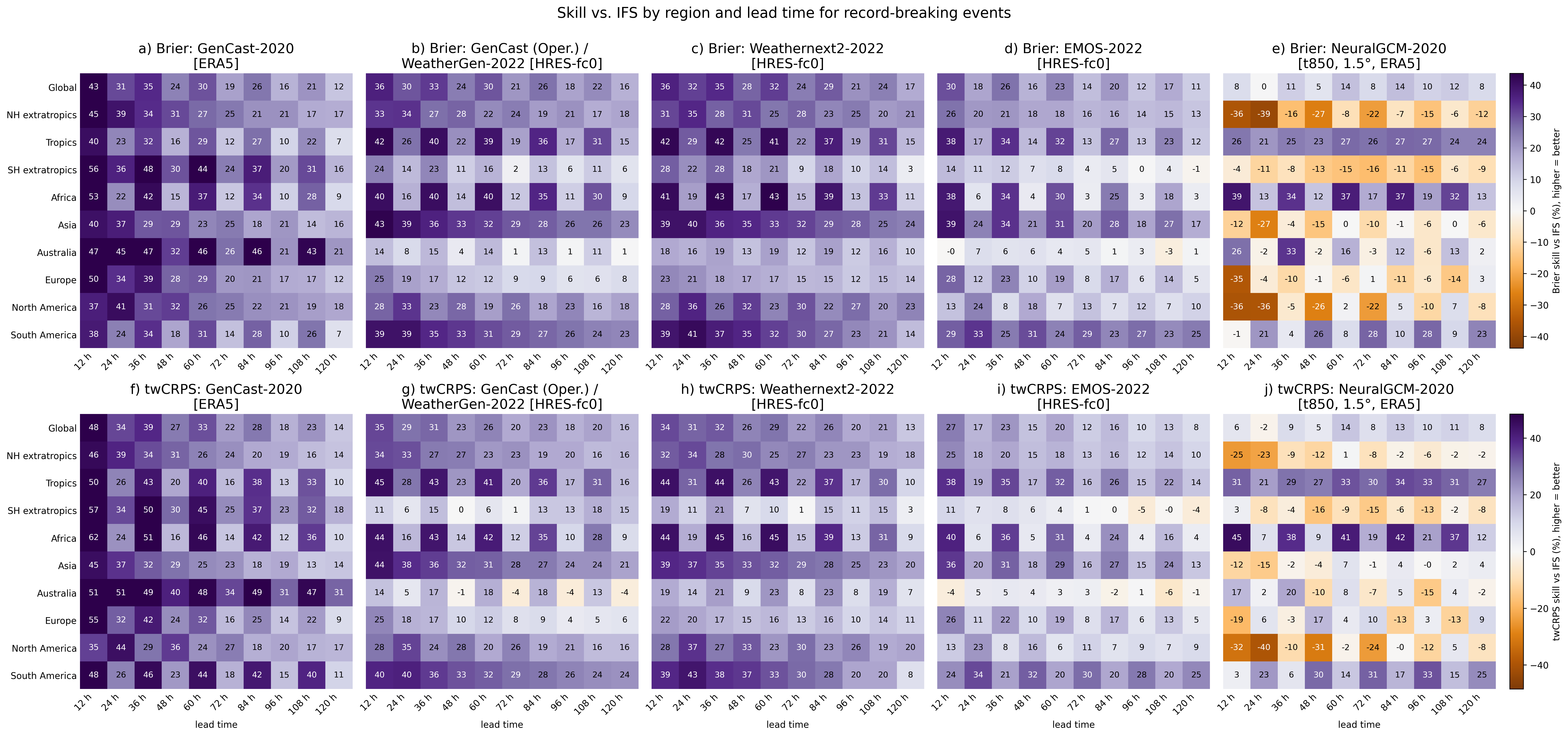}
    \caption{Skill (\%) relative to the IFS as a function of region (rows) and lead time (columns), for the Brier score (top) and twCRPS (bottom) for record-breaking heat extremes, for a,f) GenCast-2020, b,g) WeatherGen, c,h) WeatherNext2, d,i) EMOS and e,j) NeuralGCM-2020. Positive values indicate an improvement over the IFS.  Square brackets give the verification data. Note that the skill scores for the models verified against ERA5 are likely to be overoptimistic as a measure of performance as they are compared against ERA5-verified IFS.}
    \label{fig:app_skill_lead_region}
\end{figure}

\begin{figure}[b]
    \centering
    \includegraphics[width=1\linewidth]{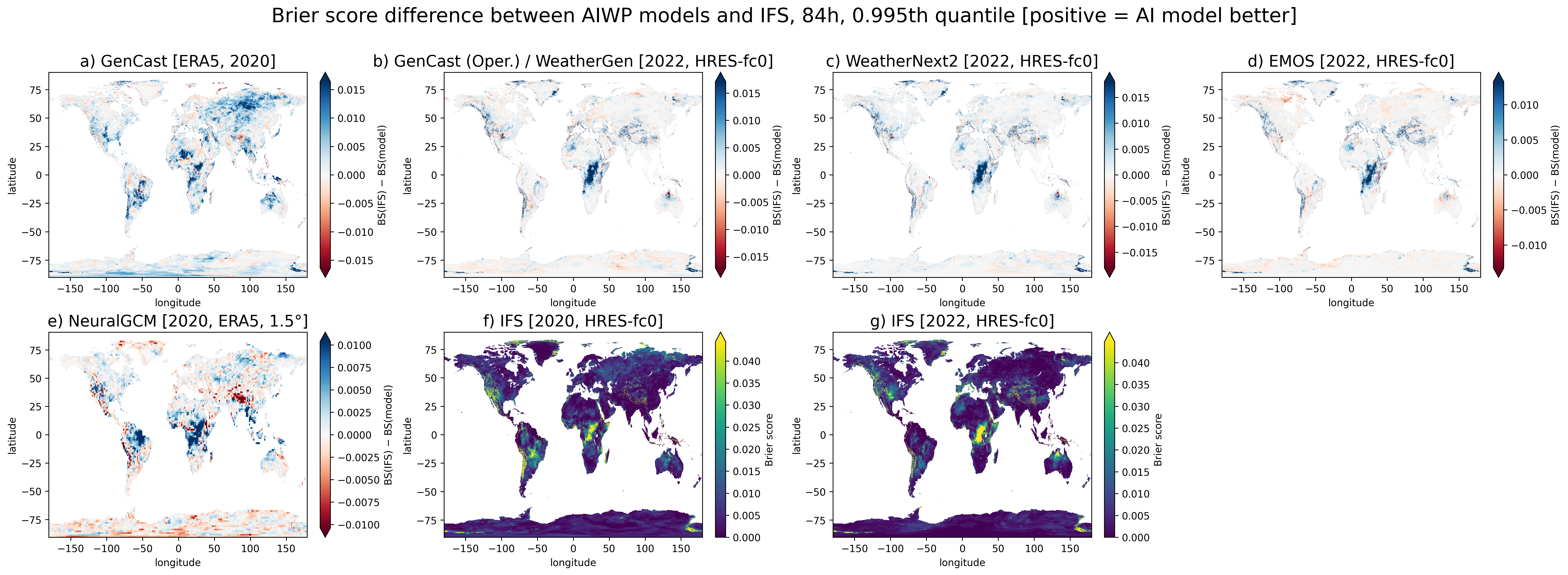}
    \caption{Difference in Brier score, BS(IFS) $-$ BS(model), at 84h lead for the 99.5th percentile, for a) GenCast-2020, b) WeatherGen, c) WeatherNext2, d) EMOS and e) NeuralGCM-2020 (t850, 1.5$^\circ$). Positive values (blue) indicate better performance than the IFS. Brier score of the IFS for f) 2020 and g) 2022.  Square brackets give the verification data. Note that the score differences for the models verified against ERA5 are likely to be overoptimistic as a measure of performance as they are compared against ERA5-verified IFS.}    
    \label{fig:app_bs_maps}
\end{figure}

\begin{figure}[b]
    \centering
    \includegraphics[width=1\linewidth]{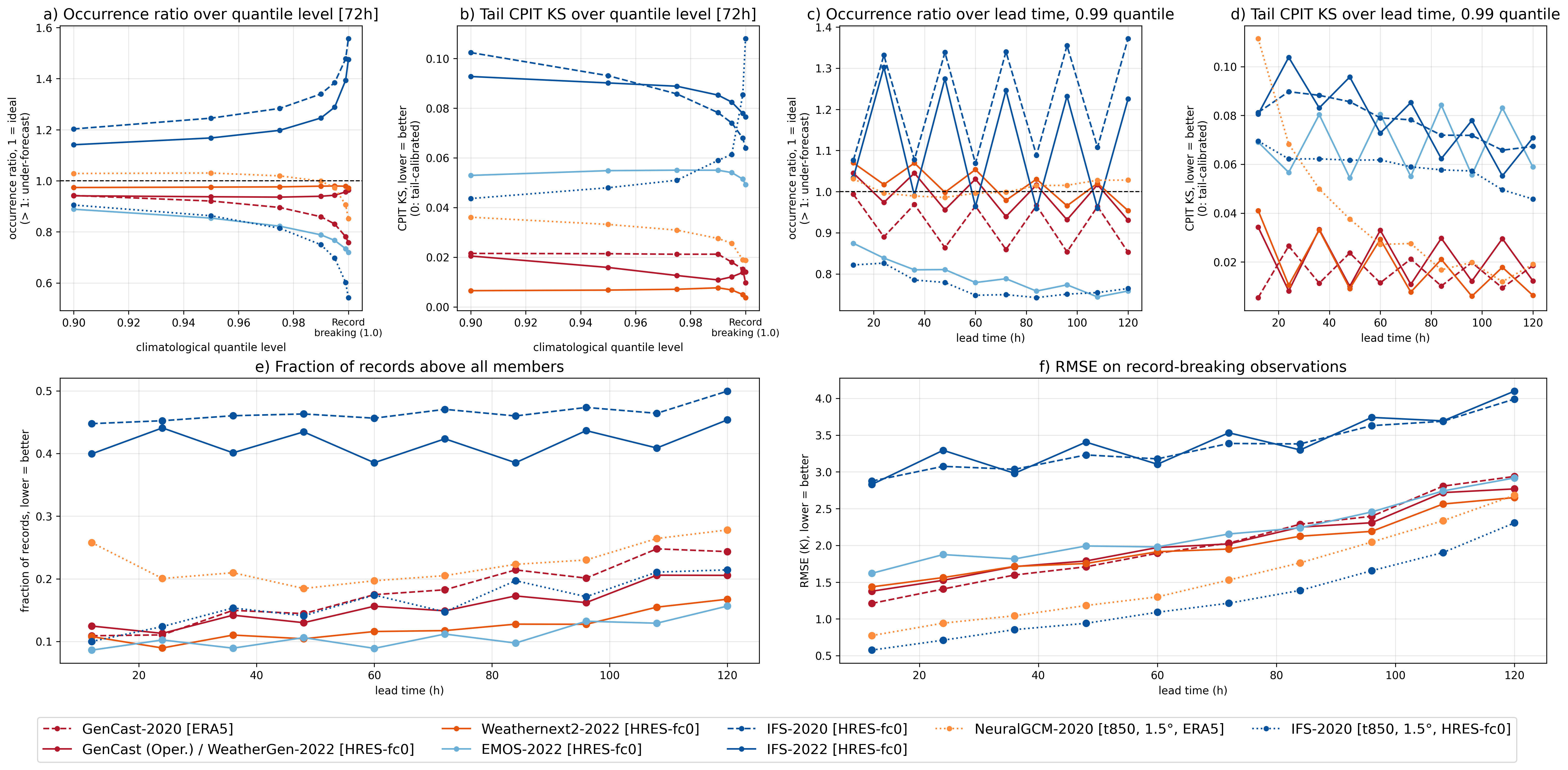}
    \caption{As Figs.~\ref{fig:extremal_scores_cali} and \ref{fig:records} for all models, including the WeatherBench2 models. a) Occurrence ratio and b) tail CPIT KS statistic at 72h as a function of the quantile level; c) occurrence ratio and d) tail CPIT KS statistic for the 99th percentile as a function of lead time; e) fraction of records above all ensemble members and f) ensemble-mean RMSE on record-breaking observations as a function of lead time. Square brackets give the verification data. Note that the skill scores for the models verified against ERA5 are likely to be overoptimistic as a measure of performance as they are compared against ERA5-verified IFS.}
    \label{fig:app_tail_calibration_records}
\end{figure}

\clearpage
\section{Evaluating against ERA5 vs. HRES-fc0}\label{app:hres_era_fcst}
It is standard to compare the IFS and operational AIWP models against HRES-fc0 and ERA5-trained models against ERA5 and different authors have argued that this is necessary for fair comparisons \citep[e.g.][]{lam_graphcast_2022, bi_accurate_2023, chen_fuxi_2023, zhang_physics-based_2026}. We also evaluate against ERA5 to test the sensitivity of our results to the choice of verification dataset. We train EMOS against ERA5, however, we also include a 2022 HRES-trained EMOS.

\begin{figure}[t]
    \centering
    \includegraphics[width=1.\linewidth]{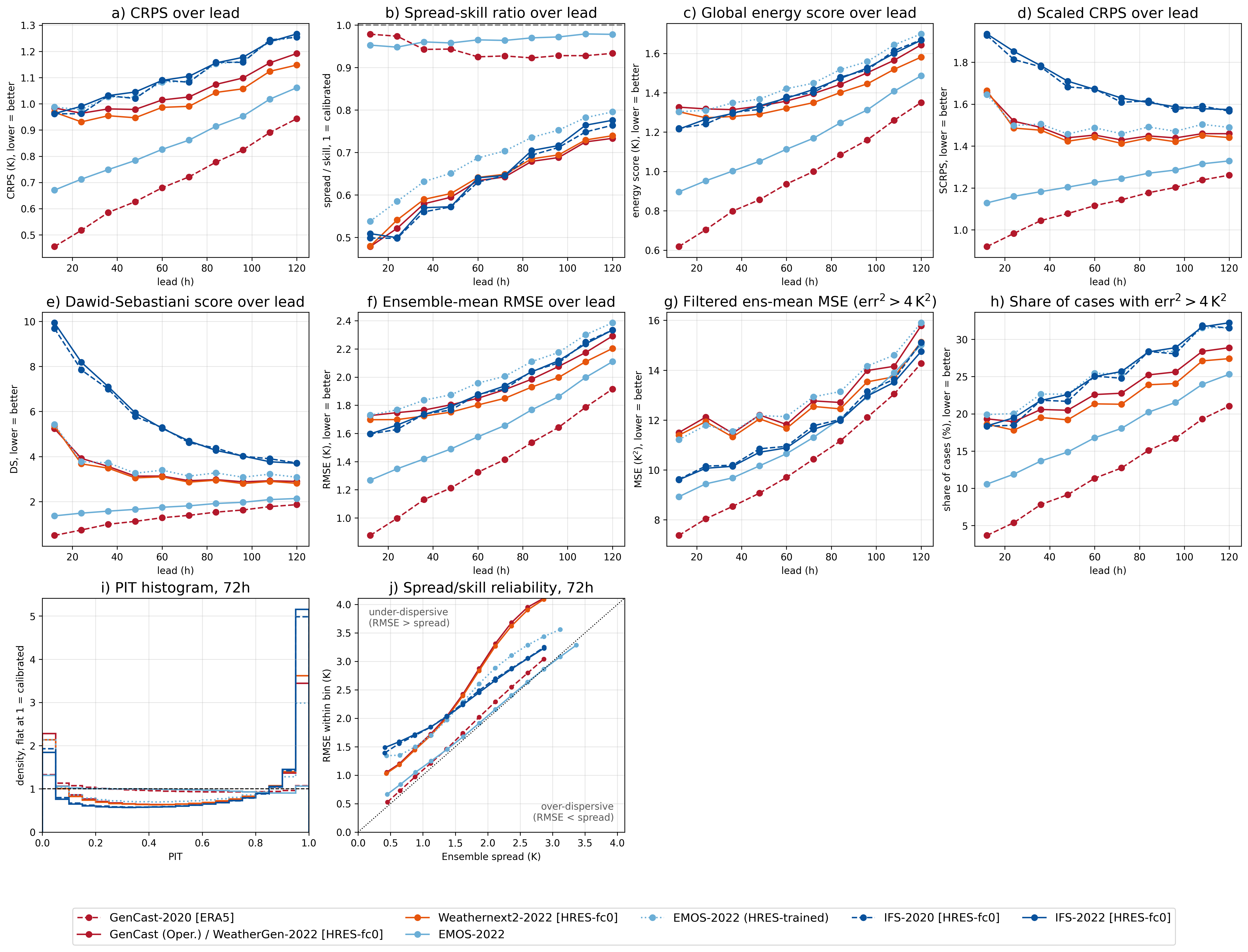}
    \caption{As Fig.~\ref{fig:app_scores_overview}, but with all models verified against ERA5. EMOS-2022 is trained on ERA5; EMOS-2022 (HRES-trained) is the HRES-fc0-trained version from the main text. Square brackets behind each model in the legend indicate the native training and verification dataset.}    \label{fig:app_overview_all_against_era5}
\end{figure}

Figure~\ref{fig:app_overview_all_against_era5} shows the overall scores when all models are verified against ERA5. GenCast (Oper.) and FGN still improve upon the IFS in terms of CRPS, but by a considerably smaller margin than when verified against HRES-fc0. The HRES-fc0-trained EMOS performs at the level of the raw IFS, whereas the ERA5-trained EMOS now outperforms both operational AIWP models. The smaller improvement of the operational AIWP models is mainly due to a deterioration in calibration: against ERA5 both models are strongly underdispersed, with spread-skill ratios close to that of the IFS. Their skill also degrades for large ensemble spread, resulting in an S-shaped spread-skill relationship. In this high-spread regime the IFS performs better, although its overall spread-skill ratio remains poor. Interestingly, GenCast (Oper.) and FGN still improve upon the IFS in terms of the scaled CRPS and the Dawid-Sebastiani score. The improvement in the Dawid-Sebastiani score is striking given their poor spread-skill ratio, since this score depends only on the first two moments of the predictive distribution. The models that are trained on and initialised from ERA5 perform much better under ERA5 verification: GenCast (ERA5) holds a large advantage over the IFS in all metrics, and the same holds, to a lesser extent, for NeuralGCM.

\begin{figure}[t]
    \centering
    \includegraphics[width=1.\linewidth]{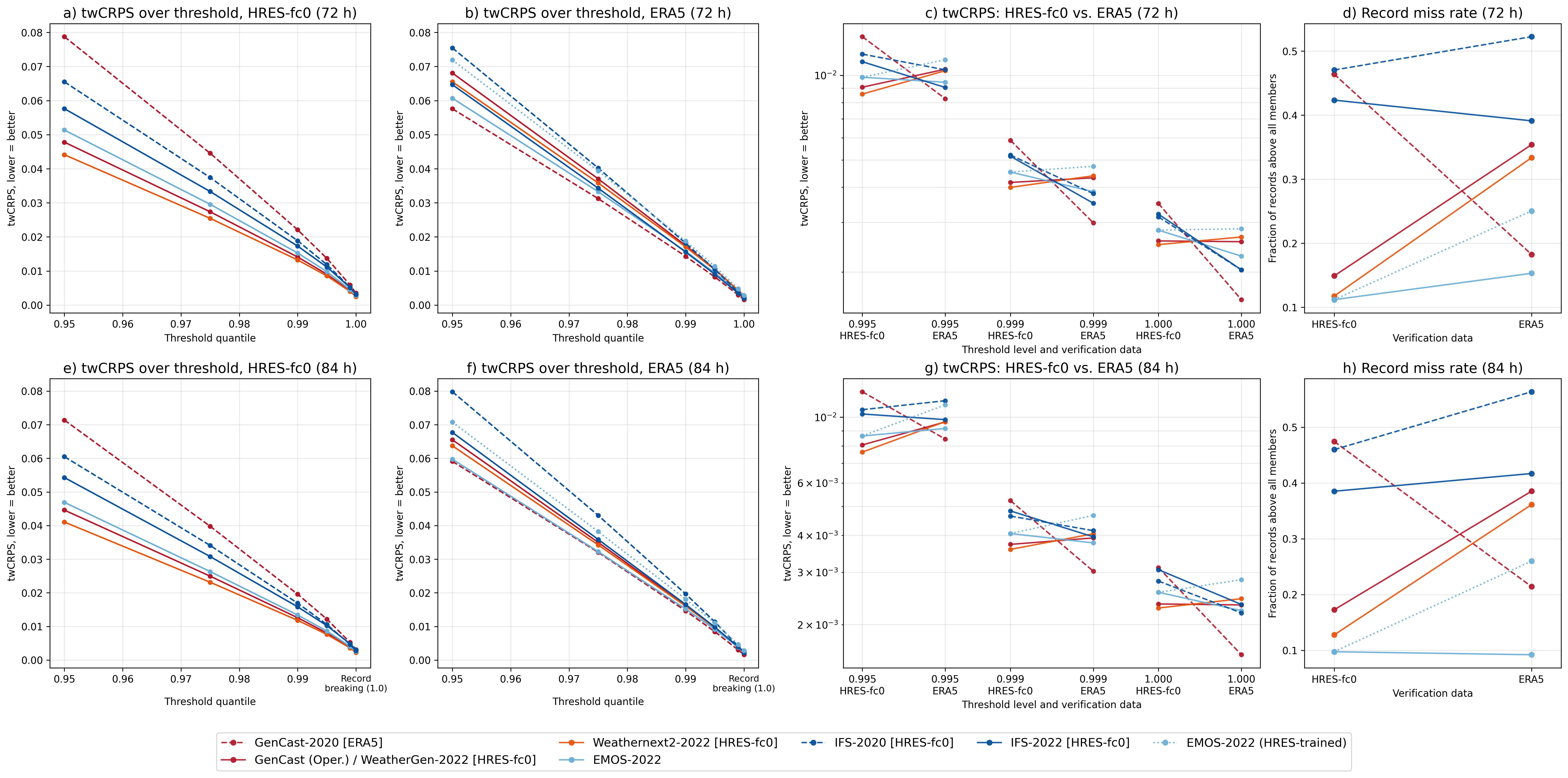}
    \caption{Extremal scores verified against HRES-fc0 and against ERA5 at 72h lead (top) and 84h lead (bottom). twCRPS as a function of the threshold quantile verified against a,e) HRES-fc0 and b,f) ERA5. c,g) twCRPS at the 99.5th and 99.9th percentiles and for records, compared between the two verification datasets. d,h) Fraction of record-breaking observations above all ensemble members (record miss rate) for both verification datasets. Square brackets behind each model in the legend indicate the native training and verification dataset.}    \label{fig:app_tail_scores_verif_target_comparison}
\end{figure}

Next we compare the performance for extremes between verification against ERA5 and against HRES-fc0. Figure~\ref{fig:app_tail_scores_verif_target_comparison} shows tail-focused proper scores together with the (non-proper) record miss rate for both verification datasets. For some thresholds, the model rankings change with the verification dataset. Although the operational AIWP models still improve upon the IFS in terms of CRPS when verified against ERA5, this improvement disappears at high thresholds. The picture is different for the ERA5-trained models that are not fine-tuned on HRES-fc0: GenCast (ERA5) strongly outperforms the IFS at all thresholds when verified against ERA5. NeuralGCM also outperforms the IFS, and interestingly does so against both verification datasets, although we note that 850\,hPa temperature is an easier forecast target than 2\,m temperature. For record-breaking heat extremes, the operational AIWP models have lower miss rates than the IFS even when verified against ERA5. GenCast (ERA5) has a miss rate similar to the IFS when verified against HRES-fc0, but a much lower one against ERA5. EMOS has the lowest miss rate under both verification datasets, consistent with its tendency to overforecast extremes.

\begin{figure}[t]
    \centering
    \includegraphics[width=1.\linewidth]{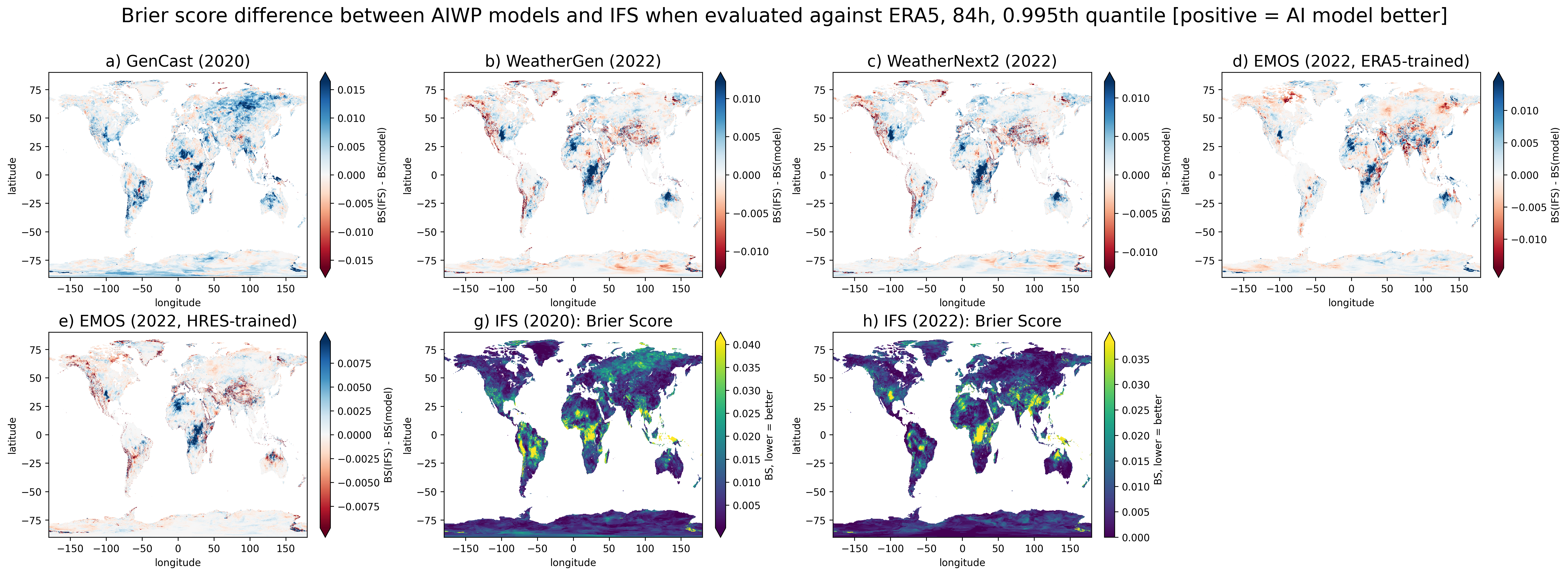}
    \caption{As Fig.~\ref{fig:app_bs_maps}, but with all models verified against ERA5 and EMOS shown both ERA5-trained (d) and HRES-fc0-trained (e). Brier score of the IFS for g) 2020 and h) 2022 against ERA5.}    
    \label{fig:app_bs_maps_all_era}
\end{figure}

Verification results for extremes therefore depend to some extent on the verification dataset. Figure~\ref{fig:app_bs_maps_all_era} shows the spatial structure of the Brier score differences against ERA5. It can be compared with Fig.~\ref{fig:app_bs_maps}, where IFS, WeatherGen and WeatherNext2 are verified against HRES-fc0. The spatial pattern of improvement is largely preserved, with particularly large improvements over Central Africa and Northern Australia. Over mountain ranges, however, the scores deteriorate. This is most likely a representativeness issue that arises from the difference in resolution between HRES-fc0 (9\,km) and ERA5 (31\,km). This difference can strongly affect data assimilation over complex terrain, so HRES-fc0 is potentially the more trustworthy reference in these regions. 

\clearpage
\section{Observation-error corrected scores}\label{sec:obserr}

In order to understand the effect of observation error between HRES-fc0 and ERA5 on the scores we use error corrected scores. Introduced by \citet{ferro_measuring_2017} these are proper scores that account for observation error and estimate the score that would be obtained against the (error-free) truth under an observation-error model. We consider two cases: 1) forecasts scored against HRES-fc0, but with ERA5 assumed to be the ground truth. 2) forecasts scored against ERA5, but with HRES-fc0 assumed to be the ground truth. The error correcting scores will then give an estimate of the true score, given a stochastic representation of the uncertainty in the verification data. This corresponds to a conservative estimate of model performance: not knowing which observational dataset is better, we represent the disagreement as a global stochastic model and incorporate that into our scoring. We note, that in practice HRES-fc0 is likely closer to the truth, not at least due to it's higher horizontal resolution and its agreement with station observations (Section~\ref{sec:verif_data}), but we see this as a complementary line of evidence for the case where we cannot say which dataset is better.

\subsection{Methods}

Let $y$ denote an observation and $x$ the underlying truth. We consider an observation model $r$. A score $s$ is \emph{everywhere unbiased} for a proper score $s_0$ under $r$ if 
$$\mathbb E_{y\mid x}\!\left[s(f,y) \mid x\right]=s_0(f,x)\quad\text{when}\quad y \mid x \sim r.$$
The observation-error corrected score $s$ then allows to estimate the underlying score $s_0$. We consider the case of binary probabilistic forecasts (of threshold exceedances). The observation error is described by the misclassification probabilities
\begin{equation}
  r_0=\Pr(y{=}1\mid x{=}0),\qquad r_1=\Pr(y{=}0\mid x{=}1)
\end{equation}
i.e. the probabilities that an event occurs in the observations (threshold exceedance) depending on whether or not it occurs in the underlying truth. \citet{ferro_measuring_2017} derive the error-corrected Brier score
\begin{equation}
  \mathrm{BS}_{\mathrm{F17}}(f,y)
  \;=\;(f-y)^2
  \;+\;\frac{r_y\,\bigl[(f-y)^2-(f-1+y)^2\bigr]}{1-r_0-r_1},
  \qquad
  r_y=\begin{cases}r_0,&y=0\\ r_1,&y=1\end{cases}.
  \label{eq:f17}
\end{equation}
However, for this score to be proper for the average of multiple forecasts it assumes that $y$ and $f$ are conditionally independent given the truth $x$, which is in general not the case if the binary event is obtained by thresholding a continuous field. Thus, we use the estimator from \citet{ramon_uncertainties_2024} in which the misclassification probabilities depending on the issued forecast, $r_0(f)$ and $r_1(f)$: \begin{equation}
  \mathrm{BS}_{\mathrm{F17\text{-}Ext}}(f,y)
  \;=\;(f-y)^2
  \;+\;\frac{r_{y}(f)\,(2f-1)(1-2y)}{1-r_0(f)-r_1(f)},
  \qquad
  r_{y}(f)=\begin{cases}r_0(f),&y=0\\ r_1(f),&y=1,\end{cases}
  \label{eq:f17ext}
\end{equation}
This is an  unbiased estimator of the against-truth Brier score, however, it is no longer proper.

We estimate the probabilities $r_0, r_1$ separately by month, but pooling all gridcells as in \citet{ramon_uncertainties_2024}. To estimate the dependence on the forecast we discretize each distribution into $B=10$ equal bins,
\begin{equation}
  \beta(f)=\min\!\bigl(\lfloor Bf\rfloor,\,B-1\bigr)\in\{0,\dots,9\}.
\end{equation}
For a quantile threshold level $q$ let $N_{q}(a,b,\beta)$ be the number of cases with truth $x=a$, verifying indicator $y=b$ and forecast bin $\beta(f)=\beta$ (pooled across the sampled lead times, initialisations, valid times in a month and grid-cells). The bin-wise estimates are the empirical conditional frequencies
\begin{equation}
  \hat r_0(\beta;q)=
  \frac{N_{q}(0,1,\beta)}{N_{q}(0,0,\beta)+N_{q}(0,1,\beta)},
  \qquad
  \hat r_1(\beta;q)=
  \frac{N_{q}(1,0,\beta)}{N_{q}(1,0,\beta)+N_{q}(1,1,\beta)} .
\end{equation}
Each case is then scored by~\eqref{eq:f17ext} with
\begin{equation}
  r_0(f)=\hat r_0\!\bigl(\beta(f);q\bigr),\qquad
  r_1(f)=\hat r_1\!\bigl(\beta(f);q\bigr).
\end{equation}
We calculate again global averages with latitude weighting and use a moving block bootstrap to quantify uncertainty. We note that estimates are unreliable if $1- r_0(f) - r_1(f) < 0.05$, in which case the score should only be taken as diagnostic \citep{ferro_measuring_2017, ramon_uncertainties_2024}, we mark these cases in the Figure. 

\subsection{Results}

\begin{figure}[h]
    \centering
    \includegraphics[width=0.8\linewidth]{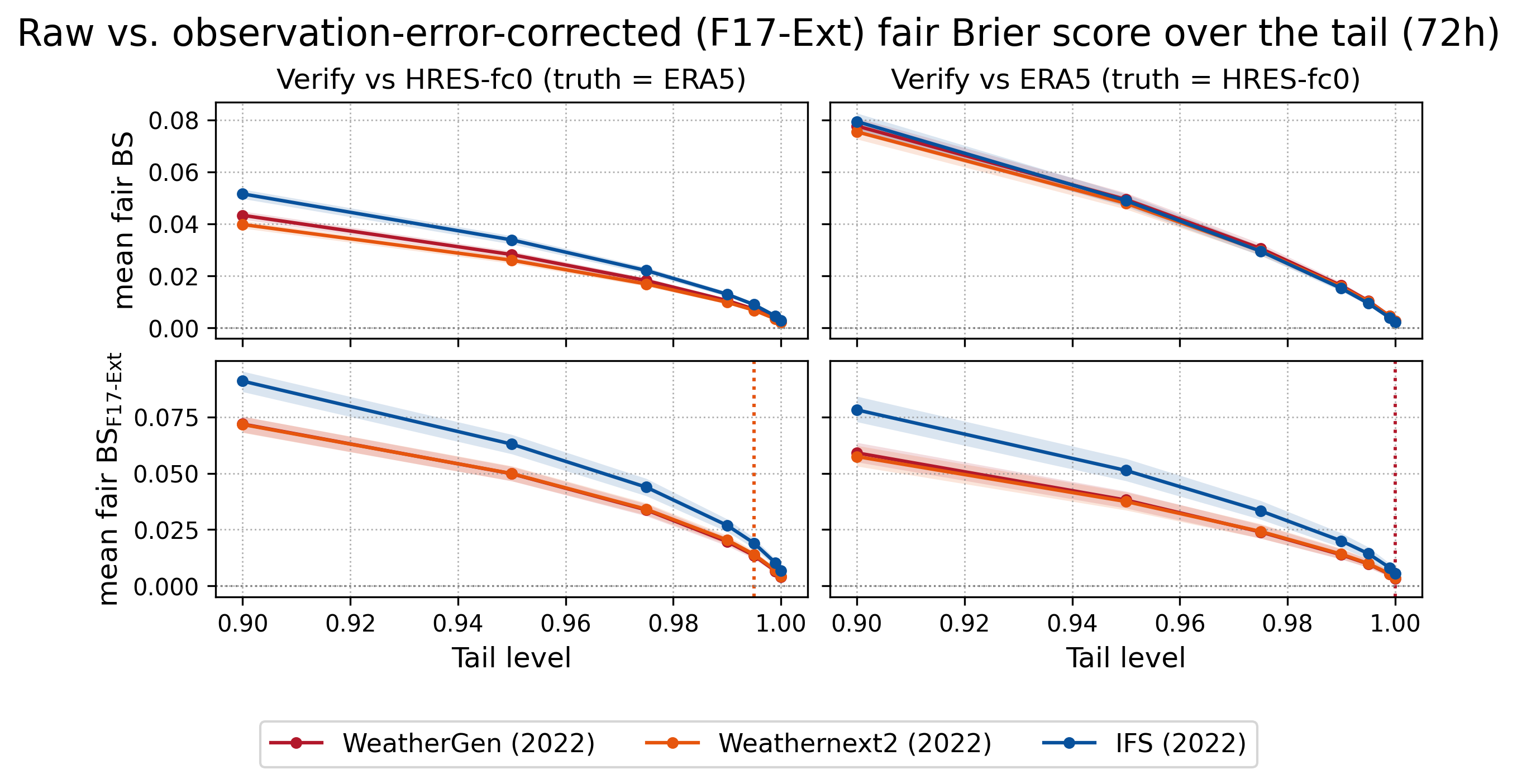}
    \caption{Raw fair Brier score (top) and observation-error-corrected fair Brier score $\mathrm{BS}_{\mathrm{F17\text{-}Ext}}$ (bottom) at 72h lead as a function of the tail level, for WeatherGen, WeatherNext2 and the IFS. Left: verified against HRES-fc0, with ERA5 as ground truth. Right: verified against ERA5, with HRES-fc0 as ground truth. Shading shows 90\% bootstrap intervals. Dotted vertical lines mark the tail level beyond which $1-r_0-r_1 < 0.05$, where the corrected score is only diagnostic \citep{ferro_measuring_2017, ramon_uncertainties_2024}.}
    \label{fig:bs_err_corrected}
\end{figure}

We estimate the disagreement between the two datasets as a stochastic misclassification process, with misclassification probabilities $r_0(f)$ and $r_1(f)$ estimated per month and forecast-probability bin. We compute the classical fair Brier score for AIWP and IFS models, as well as the error corrected score, which is then an estimate of the expected Brier score against the assumed truth under this error process (Fig~\ref{fig:bs_err_corrected}). 

The classical uncorrected scores show for ERA5 verification near-equal performance at moderate thresholds and an IFS advantage at the highest ones, whilst for HRES-fc0 verification the AIWP models outperform. The corrected scores, however, differ and both for HRES-fc0 verification and ERA5 verification they give a consistent picture of AIWP models outperforming the IFS. This means that under a stochastic representation of the uncertainty in the verification data, the AIWP advantage for extremes therefore holds for both datasets, heavily suggesting that even with a conservative estimate of skill, accounting for observational uncertainty, AIWP models do outperform the IFS.

\clearpage
\bibliographystyle{abbrvnat}
\bibliography{references}
\end{document}